\documentclass[12pt,draftclsnofoot,onecolumn]{IEEEtran}
\usepackage{comment}
\usepackage{amsmath,amssymb,bm}
\usepackage{tabularx}
\usepackage{graphicx}
\usepackage{booktabs} 
\usepackage{caption}
\usepackage{subcaption}
\usepackage{xcolor}
\usepackage{cite}
\usepackage[linesnumbered,ruled,vlined]{algorithm2e}
\newcommand\norm[1]{\left\lVert#1\right\rVert}

\newcommand{\tran}{^{\mbox{\scriptsize T}}}
\newcommand{\herm}{^{\mbox{\scriptsize H}}}

\DeclareMathOperator{\Tr}{Tr}

\makeatletter
\def\thickhline{%
  \noalign{\ifnum0=`}\fi\hrule \@height \thickarrayrulewidth \futurelet
   \reserved@a\@xthickhline}
\def\@xthickhline{\ifx\reserved@a\thickhline
               \vskip\doublerulesep
               \vskip-\thickarrayrulewidth
             \fi
      \ifnum0=`{\fi}}
\makeatother

\newlength{\thickarrayrulewidth}
\ifCLASSINFOpdf
\else
\fi
\begin{document}
\bstctlcite{IEEEexample:BSTcontrol}
% paper title
% Titles are generally capitalized except for words such as a, an, and, as,
% at, but, by, for, in, nor, of, on, or, the, to and up, which are usually
% not capitalized unless they are the first or last word of the title.
% Linebreaks \\ can be used within to get better formatting as desired.
% Do not put math or special symbols in the title.
%\textcolor{red}{Joint Coherent and Non-coherent Signal Processing in a Heterogeneous Network} 
\title{Joint Coherent and Non-Coherent Detection and Decoding Techniques for Heterogeneous Networks}
% Joint Coherent and Non-Coherent Decoding Techniques for Heterogeneous Networks
%
%
% author names and IEEE memberships
% note positions of commas and nonbreaking spaces ( ~ ) LaTeX will not break
% a structure at a ~ so this keeps an author's name from being broken across
% two lines.
% use \thanks{} to gain access to the first footnote area
% a separate \thanks must be used for each paragraph as LaTeX2e's \thanks
% was not built to handle multiple paragraphs
%

%\author{Leatile Marata, Onel Luis Alcaraz López, Hamza Djelouat, Markus Leinonen, and 
%}
 % 6G Flagship (grant 346208) in part by  6G Flagship (grant 346208) and in part by the Academy of Finland (grant 319485)

\author{
\IEEEauthorblockN{Leatile Marata, Onel Luis Alcaraz López, \textit{Member, IEEE}, Hamza Djelouat, \textit{Student Member, IEEE}, Markus Leinonen, \textit{Member, IEEE}, Hirley Alves, \textit{Member, IEEE}, and Markku Juntti, \textit{Fellow, IEEE}}
\thanks{The authors are with Centre for Wireless Communications -- Radio Technologies, FI-90014, University of Oulu, Finland. e-mail: \{leatile.marata, onel.alcarazlopez, hamza.djelouat, markus.leinonen.fi, hirley.alves, markku.juntti\}@oulu.fi.}
\thanks{This work is supported by Academy of Finland (Grants n.346208 (6G Flagship) and  n.340171,). The work of Leatile Marata was supported in part by the Riitta ja Jorma J. Takanen Foundation and the Botswana International University of Science and Technology. The work of Onel López was supported in part by the Finnish Foundation for Technology Promotion. The work of Hamza Djelouat was supported in part by the Tauno Tönning Foundation, the Riitta ja Jorma J. Takanen Foundation, and the Nokia Foundation. }
}

\maketitle

% As a general rule, do not put math, special symbols or citations
% in the abstract or keywords.
%\textcolor{red}{Even so, most of these solutions are limited to a scenario where MTDs operate alone in the assigned spectrum resources, which limits their potential application in 5G and beyond (5GB) cellular networks, where harmonious coexistence of heterogeneous services may be required.}
%\vspace{-10mm} %%COMPRESSION TWEAK
%\begin{spacing}{1.33} %%COMPRESSION TWEAK
\begin{abstract} 
Cellular networks that are traditionally designed for human-type communication (HTC) have the potential to provide cost effective connectivity to machine-type communication (MTC). However, MTC is characterized by unprecedented traffic in cellular networks, thus posing a challenge to its successful incorporation. In this work, we propose a unified framework for amicable coexistence of MTC and HTC. We consider a heterogeneous network where machine-type devices coexist with enhanced mobile broadband (eMBB) devices and propose transceiver techniques that promote efficient signal recovery from these devices. For this, we present an eMBB pilot and MTC data generation strategy that facilitates joint coherent decoding of eMBB data and non-coherent decoding of MTC data. Furthermore, we assess the feasibility of coexistence using receiver operating characteristics, outage probability, and normalized mean square error (NMSE). Our numerical results reveal that a harmonious coexistence of the heterogeneous services can be achieved with properly configured average signal-to-noise ratios and pilot length. 
%\textcolor{red}{Our numerical analysis evaluates the coexistence behavior of eMBB and MTC using receiver operating characteristics, outage probability, and normalized mean square error (NMSE), which reveal that harmonious coexistence of the heterogeneous services can be achieved with properly configured average received signal-to-noise ratio and pilot length. }

%%COMPRESSION TWEAK
\textbf{Index Terms:} Coherent and non-coherent decoding, detection, enhanced mobile broadband, grant-free,  
machine-type communication, sparse signal recovery.
\end{abstract}
%\end{spacing}
%\vspace{-5mm}%%COMPRESSION TWEAK

%\begin{IEEEkeywords} Coherent and non-coherent decoding, detection, enhanced mobile broadband, grant-free,  machine-type communication, sparse signal recovery. 
%\end{IEEEkeywords}

% For peer review papers, you can put extra information on the cover
% page as needed:
% \ifCLASSOPTIONpeerreview
% \begin{center} \bfseries EDICS Category: 3-BBND \end{center}
% \fi
%
% For peerreview papers, this IEEEtran command inserts a page break and
% creates the second title. It will be ignored for other modes.
\IEEEpeerreviewmaketitle

\section{Introduction}
% The very first letter is a 2 line initial drop letter followed
% by the rest of the first word in caps.
% 
% form to use if the first word consists of a single letter:
% \IEEEPARstart{A}{demo} file is ....
% 
% form to use if you need the single drop letter followed by
% normal text (unknown if ever used by the IEEE):
% \IEEEPARstart{A}{}demo file is ....
% 
% Some journals put the first two words in caps:
% \IEEEPARstart{T}{his demo} file is ....
% 
% Here we have the typical use of a "T" for an initial drop letter
% and "HIS" in caps to complete the first word.

%% INTRODUCTION TO MTC CHARACTERISTICS 
\IEEEPARstart{M}{achine}-type communication (MTC) is of paramount importance in the realization of a digitally connected society due to its application in cellular Internet of Things (IoT) \cite{liu2018massive,liu2018massive2,mahmood2021machine}. As a result, there is an accelerated deployment of machine-type devices (MTDs) that perform tasks such as health, pollution, energy consumption, and infrastructure monitoring \cite{ghavimi2014m2m,zanella2014internet}. To provide cost-effective connectivity to these devices, it is advantageous to utilize existing cellular networks \cite{shariatmadari2015machine,sharma2019toward}. However, cellular connectivity is traditionally designed for human-type communication (HTC) so that the key differences hinder seamless HTC/MTC integration. For instance, contrary to HTC, MTC is characterized by sporadic uplink traffic of short packets \cite{liu2018massive2}. Furthermore, the massive number of MTDs in a cell makes it infeasible to allocate orthogonal pilot sequences, thus, challenging conventional multiple access transceiver techniques \cite{senel2018grant,liu2018sparse}.

%% PROPOSED SOLUTIONS TO THE GRANT FREE 
To handle MTC's sporadic traffic and massiveness of MTDs, grant-free non-orthogonal multiple access (GF-NOMA) techniques have been proposed \cite{shahab2020grant,popovski2019wireless}. These techniques allow the MTDs to transmit without waiting for a permission, which is conventionally granted through a handshaking process \cite{senel2018grant}. However, collisions become unavoidable, thus, potentially degrading the system performance. To resolve the collisions, several signal detection algorithms which exploit the sporadic traffic of MTDs have recently  been proposed \cite{senel2018grant,liu2018sparse,islam2016power,zhang2011sparse,huang2022noncoherent}. In essence, the received signal is a compressed measurement of sparse (effective) channels for different users, while the base station (BS) is tasked with identifying and estimating these channels. This two-fold inference task involving sparse signal recovery in MTC gives rise to the problem of compressed sensing based multi-user detection (CS-MUD), which can be formulated and solved using compressed sensing theory \cite{Candes-Romberg-Tao-06,donoho2006compressed,Haupt-Nowak-06}. Specifically, CS-MUD problems can be posed as coherent or non-coherent signal recovery problems \cite{lee2017packet,lancho2021joint}. Coherent recovery uses channel state information (CSI), hence, it is only efficient in scenarios where the payload is much greater than the metadata \cite{ahn2019ep,wei2018joint}. On the other hand, non-coherent recovery is carried out without CSI and it is only favorable for short packet data traffic \cite{chen2019covariance,huang2021compressed,huang2020design,chowdhury2016scaling}.

From the preceding discussions, it is apparent that GF-NOMA suited for MTDs differs from grant based medium access techniques used for HTC \cite{miao2016fundamentals}. Fundamentally, GF-NOMA reduces the access latency at the cost of increasing collisions, while the opposite is true for grant based medium access. In addition, while coherent decoding techniques are suitable for eMBB transmissions, non-coherent decoding techniques are specifically well suited for MTC. In spite of these differences, a 5G  and  beyond  (5GB) cellular network must accommodate both eMBB and MTC using limited resources \cite{senel2018human}. It is therefore important to note that joint coherent decoding of eMBB devices and non-coherent detection of MTDs has great potential to alleviate the spectrum scarcity problem by promoting harmonious { coexistence of the different services}. To this end, the main focus of our paper is to introduce a framework for coexistence of MTC and HTC services (i.e., eMBB) within the same spectrum resource block. We subsequently present a brief literature review of some of the works that gave us the impetus to pursue this work.

\subsection{Related Literature}
Sparse signal recovery (SSR) algorithms have for long been applicable in signal processing frameworks, where signals have sparse representations in a certain basis \cite{donoho2006compressed,choi2017compressed}. In light of this, the data traffic from MTDs can be captured using sparse representations. Furthermore, the number of MTDs is expected to grow to  around $10^6$ devices per cell in the next few years, among which only a limited set will concurrently be active \cite{di2021dynamic}. From the receiver perspectives, further deployments of MTDs result in a complex SSR problem, which may cause performance degradation. Owing to this, there is an increased research interest in CS-MUD \cite{di2020detection,zhang2021unifying}.

Among some of the markedly proposed solutions to CS-MUD problems are graphical models due to their ability to represent complex information using probability distributions \cite{kschischang2001factor,wainwright2008graphical}. Based on this property, message passing (MP) algorithms on graphical models have been extensively used to handle the sporadic traffic from MTDs, e.g., using belief propagation (BP) \cite{wang2018framework}, approximate MP (AMP) \cite{wei2018joint,rush2018finite,takeuchi2020convolutional,rangan2019vector,jiang2022performance}, and expectation propagation (EP) \cite{ahn2019ep}.

Wei \textit{et al.} \cite{wei2018joint} proposed a novel minimum mean square error (MMSE) denoiser for AMP in massive deployments of MTDs. Their work showed that it is possible to drive both the probability of miss detection and the probability of false alarm to zero in massive multiple-input multiple-output (MIMO) setups. Their algorithm is designed for coherent detection, hence does not exploit the short packet structure of MTDs. Senel \textit{et al.}  \cite{senel2018grant} introduced a non-coherent AMP algorithm that uses a likelihood ratio of the information symbols. Different from \cite{senel2018grant}, where the assumed prior distributions do not take activity pattern of MTDs into consideration, Tang \textit{et al.} proposed a Bayes optimal algorithm in \cite{tang2020device}. Rodrigo \textit{et al.} \cite{di2021dynamic} analyzed a new dimension to device activity detection by considering asynchronous device activity in a given coherence interval. Another perspective of MP is seen in \cite{ahn2019ep}, where Ahn \textit{et al.} proposed a joint user activity detection and channel estimation algorithm based on EP. In spite of its ability to handle any activity pattern, a key limitation lies in the assumption of a single-antenna BS scenario, which is not practical for supporting massive connectivity. Huang \textit{et al.} \cite{huang2022noncoherent} proposed generalised AMP and deep learning based non-coherent detection algorithms for scenarios subject to Rician fading, where the activity pattern differs from one MTD to another. Furthermore, Huang \textit{et al.}  \cite{huang2020design} proposed data-driven non-coherent transceiver techniques for short packet transmissions. Apart from MP based approaches, other solutions to CS-MUD problem exploit convex optimization techniques. To this end, Chen  \textit{et al.} \cite{chen2019covariance} proposed a covariance aided non-coherent detection algorithm, and Djelouat \textit{et al.} \cite{djelouat2021joint} proposed a joint user identification and channel estimation  algorithm using alternating direction method of multipliers (ADMM). 

One shortcoming of the previous works is the assumption that MTDs operate alone in the corresponding resource block. This may disable their application to 5GB cellular networks where heterogeneous services, i.e., ultra-reliable low latency communication (URLLC), MTC, and eMBB may need to coexist amicably \cite{popovski2019wireless,senel2018human,alsenwi2019embb,abedin2018fog}. As previously mentioned, existing HTC cellular networks are appealing for providing connectivity to MTC. However, successful incorporation of MTC into cellular networks designed for HTC requires tailoring algorithms that enable i) resolution of multiple access interference and collisions of MTDs, ii) efficient short packet data transmissions, and iii) coexistence of MTC with other services.

In our previous work \cite{marata2021joint}, we proposed a pilot design strategy that enables efficient channel estimation of a single eMBB device in the presence of MTC. In addition, we presented detection of MTDs with successive interference cancellation (SIC) of eMBB data. In this regard, \cite{marata2021joint} was a first necessary step for enabling MTC-eMBB coexistence by studying the single eMBB user scenario and the first phase of the transmission block, i.e., metadata/pilot training phase. We had assumed that the interference power on the detection process of MTDs emanated exclusively from the receiver noise and the channel estimation error of the eMBB device. Meanwhile, in this paper, we advance the state-of-the-art by considering a more general multi-eMBB user setup and the entire transmission block, i.e., metadata and data decoding. Under this setup, signals cannot be guaranteed to be orthogonal to each other in the second transmission block. This results in stronger interference and thus, the present work provides a more practical coexistence solution for the eMBB and MTC services.  To the best of our knowledge, this is the first work that presents a unified framework to accommodate joint coherent and non-coherent signal reception in a heterogeneous 5GB network consisting of eMBB and MTC traffic.

\subsection{Contributions}
We consider a heterogeneous network, where eMBB devices and MTDs are co-hosted in the same spectrum resource block. By acknowledging that the two services have heterogeneous requirements and characteristics, we propose service-specific techniques: coherent decoding for eMBB devices and non-coherent decoding for MTDs. Coherent decoding is facilitated by properly designed pilot symbols for eMBB devices, while non-coherent decoding is achieved by an optimized message generation strategy for MTDs. Furthermore, we consider joint activity and data decoding (JADD) of MTDs as a CS-MUD problem. This CS-MUD problem is more challenging than those in, e.g., \cite{wei2018joint,ahn2019ep,djelouat2021joint}, due to possible interference from eMBB signals. 

The key contributions of our paper are:
\begin{itemize}
    \item We propose novel transceiver signal processing techniques that facilitate the joint detection/decoding of eMBB and MTC data within the same spectrum resource block. This facilitates coexistence of heterogeneous services, thus, addressing a practical need for 5GB cellular connectivity. 
    \item We propose a novel message generation strategy for MTDs. The generated messages are the transmitted symbols from MTDs and also used as a measurement (or sensing) matrix in compressed sensing terminology \cite{donoho2006compressed}. This matrix is designed such that part of its structure maintains orthogonality with the pilot sequences of the eMBB devices. The orthogonality makes it possible to get interference-free channel estimates, which leads to more efficient coherent decoding and SIC.
    \item We adapt state-of-the-art MTC detection algorithms to the considered heterogeneous scenario. Specifically, AMP \cite{wei2018joint,senel2018grant}, $\ell_{2,1}$  minimization \cite{boyd2011distributed,boyd2004convex}, expectation-maximization sparse Bayesian learning (EM-SBL) \cite{zhang2011sparse}, and simultaneous orthogonal matching pursuit (SOMP) 
\cite{cai2011orthogonal} algorithms are applied to recover the transmitted messages of MTDs using the received signal remaining after SIC of the eMBB devices.  
    \item Finally, we provide a numerical performance analysis for the coexistence of the eMBB and MTC traffic. Our findings reveal that harmonious coexistence is viable when the operating point of the heterogeneous network is appropriately adjusted by tuning the average signal-to-noise ratio (SNR) and pilot length parameters.  
\end{itemize}
\subsection{Organization and Notation}
The remainder of this paper is organised as follows. In Section~\ref{formulate}, we describe the system model. In Section \ref{detections}, we present the coherent data processing of the eMBB data, while in Section~\ref{JADD}, we introduce the problem of non-coherent detection of MTDs. Section~\ref{SSRALG} deals with SSR algorithms for JADD. Section~\ref{results} presents numerical results, and lastly, in Section~\ref{conclude}, we draw the conclusions and recommend possible research directions. 
%
%\textcolor{red}{,while $\Bar{\mathbf{a}}_i$ is the $i$-th row of matrix $\textbf{A}$}
\par \textbf{Notation:} Boldface lowercase and boldface uppercase letters denote column vectors and matrices, respectively. Moreover, $\mathbf{a}_i$ and $a_{i,j}$ are the $i$-th column and the element in the $i$-row, $j$-th column of matrix $\mathbf{A}$, respectively, while $a_i$ is the $i$-the element of vector $\mathbf{a}$. The superscripts $(\cdot)^*$, $(\cdot)\tran$, and $(\cdot)\herm$ denote the conjugate,  transpose, and conjugate transpose operations, respectively. We denote the circularly symmetric complex Gaussian distribution with mean $\mathbf{a}$ and covariance $\mathbf{B}$ by $\mathcal{CN}(\mathbf{a},\mathbf{B})$ while $\mathbb{E}\{\cdot\}$ is the expectation operator. If $\mathbf{D}_1,\cdots,\mathbf{D}_N$ are square matrices, then  $\mathbf{D}=\mathrm{diag}(\mathbf{D}_1,\mathbf{D}_2,\cdots, \mathbf{D}_{N})$ creates a matrix whose block diagonal  matrices are $\mathbf{D}_1,\mathbf{D}_2,\cdots, \mathbf{D}_{N}$, while $\mathrm{diag}\{a_{1},a_{2},\cdots,a_{n}\}$ creates a diagonal matrix whose main diagonal terms are $a_{1},a_{2},\cdots,a_{n}$. For both matrices and vectors, the hat notation indicates an estimate, e.g., $\hat{x}$ is the estimate of $x$;  $\mathbb{C}$ and  $ \mathbb{R}$ refer to complex and real domains, respectively. Finally, $\norm{\cdot}_{F}$ and $\norm{.}_{p}$ denote the Frobenius and $\ell_p$ norms, respectively and $\binom{a}{b}$ is the binomial coefficient. Table~\ref{Acronyms} summarizes the main acronyms and symbols used in this paper.
 \begin{table*}
    \centering
  \caption{Acronyms and main symbols}
 \begin{tabular}{p{1.1cm}|p{7.4cm}|p{1 cm}|p{6.9cm}}
\thickhline
5GB &   5G and beyond &  NMSE & normalized mean square error   \\ 
eMBB &   enhanced mobile broadband &  MTDs & machine-type communication devices   \\ 
BS &   base station &  NMSE & normalized mean square error   \\ 
AMP &  approximate message passing &  SBL & sparse Bayesian learning   \\
SOMP &  simultaneous orthogonal matching pursuit  &  ADMM & alternating direction method of multipliers   \\ 
JADD &  joint activity detection and decoding  &  SIC & successive interference cancellation   \\ 
CS-MUD &  compressed sensing multiuser detection  &  NOMA & non orthogonal multiple access   \\ 
PMD &  probability of miss detection  &  PFA & probability of false alarm   \\EM &  expectation-maximization  &  MTC & machine-type communication  \\ 
\hline
$\mathcal{N}, N$ & set and number of MTDs & $p_n^{\mathrm{UL}}$ & uplink transmission power of the $n$-th MTD   \\ 
$\mathcal{E}, E$ & set and number of eMBBs & $\rho_e^{\mathrm{UL}}$ & uplink transmission power of the $e$-eMBB device   \\
$\gamma_n$ & large scale fading coefficients of the $n$-th MTD & $\beta_e$ & large scale fading coefficients of the $e$-th eMBB device   \\ 
$\Gamma_e$ & signal to interference plus noise ratio for the $e$-th eMBB device & $M$ & the number of antennas  \\ 
$\bm{\omega}_e$ & receive combining weights & $b$ & the number of bits   \\
$T$ & coherence interval & $L$ & pilot length  \\
$s_e[k]$ & $k$-th symbol transmitted by $e$-th eMBB device & $s_n[k]$ & $k$-th symbol transmitted by  $n$-th MTD  \\
$t_{max}$ & maximum allowable number of iterations & $K$ &the number of active MTDs  \\
$r$ & data rate & $Q$ &the number of sequences allocated to each MTD \\
$\epsilon$ & average activation probability of the MTDs & $P_{out}$ & average outage probability of the eMBB devices \\
\thickhline
\end{tabular}
\label{Acronyms}
\end{table*}

\section{System Model}\label{formulate}
%% I HAVE CHANGED THE NOTATION OF THE POWER FOR MTDS AND ALSO INTRODUCED A NEW CHANNEL VARIABLE. I ALSO HIGHLIGHT THAT I DID NOT SEE THE NEED TO DEFINE A SET FOR EMBBS 
  We consider a heterogeneous network depicted by Fig.~1(a), where a BS with an $M$-element array serves a set $\mathcal{N} = \{1,\cdots,N\}$ of single-antenna MTDs and a set $\mathcal{E} = \{1,\cdots,E\}$ of eMBB devices. The uplink transmission powers are $\rho_e^{\mathrm{UL}}$, ${e \in \mathcal{E}} $, for the $e$-th eMBB device and $p_n^{\mathrm{UL}}$, $n \in \mathcal{N}$, for the $n$-th MTD. We assume that all $E$ eMBB devices are active in a coherence interval of $T$ symbols, while only $K$ out of the $N$ MTDs are active. The justification for this assumption is that eMBB devices are generally active for longer periods of time and usually apply scheduled access as opposed to MTDs, which only transmit short data messages with sporadic activation \cite{bairagi2020coexistence}. The set of active MTDs, denoted by ${\mathcal{K}\subseteq{\mathcal{N}}}$, is unknown to the BS. Moreover, the MTDs are assumed to activate with probability $\epsilon$ in each coherence interval, hence, on average $\epsilon N$  MTDs are active in each coherence interval.
The channel between the $e$-th eMBB device and the BS is modelled as $\mathbf{h}_e\sim\mathcal{CN}( 0,\beta_e\mathbf{I}) \in \mathbb{C}^{M\times 1}$ and between the $n$-th MTD and the BS as  $\tilde{\mathbf{g}}_n\sim\mathcal{CN}( 0,\gamma_n\mathbf{I}) \in \mathbb{C}^{M\times 1}$, where $\{\beta_{e}\}$ and $\{\gamma_{n}\}$ are the large scale fading coefficients. All devices are assumed to be stationary and, therefore, the BS knows $\{\beta_{e}\}$ and $\{\gamma_{n}\}$.\begin{figure}[t]
 \centering
\begin{subfigure}{0.45\textwidth}
  \centering
    \includegraphics[scale=0.15,width =\textwidth]{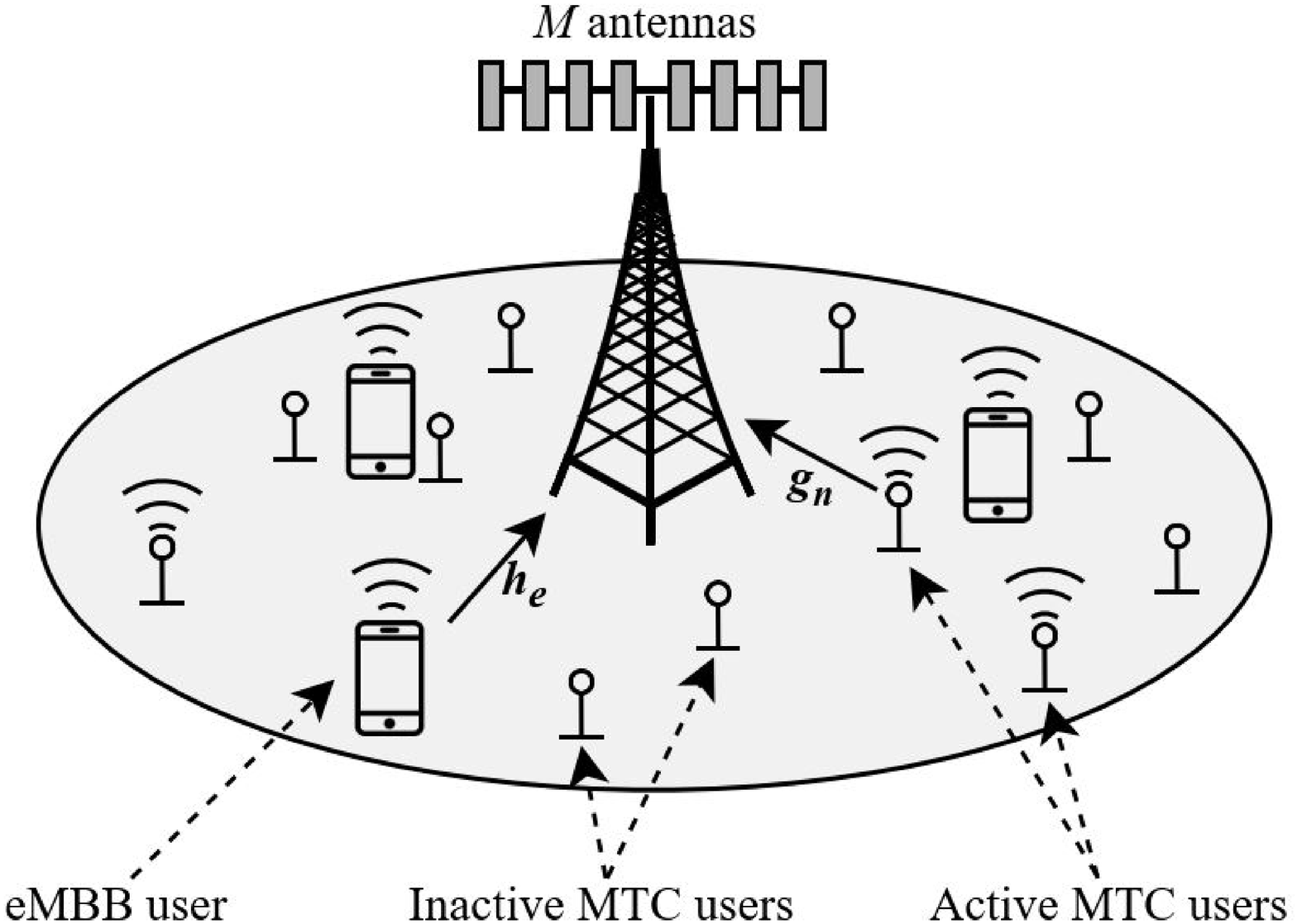}
   % \caption{A heterogeneous network with \textcolor{blue}{$E$} eMBB devices and a massive number $N$ of MTDs, from which $K$ are active \textcolor{blue}{MTDs} and $N-K$ are inactive.}
    \label{figMTC}
\end{subfigure}
\
\begin{subfigure}{0.45\textwidth}
  \centering
   \includegraphics[scale=0.15,width = \textwidth]{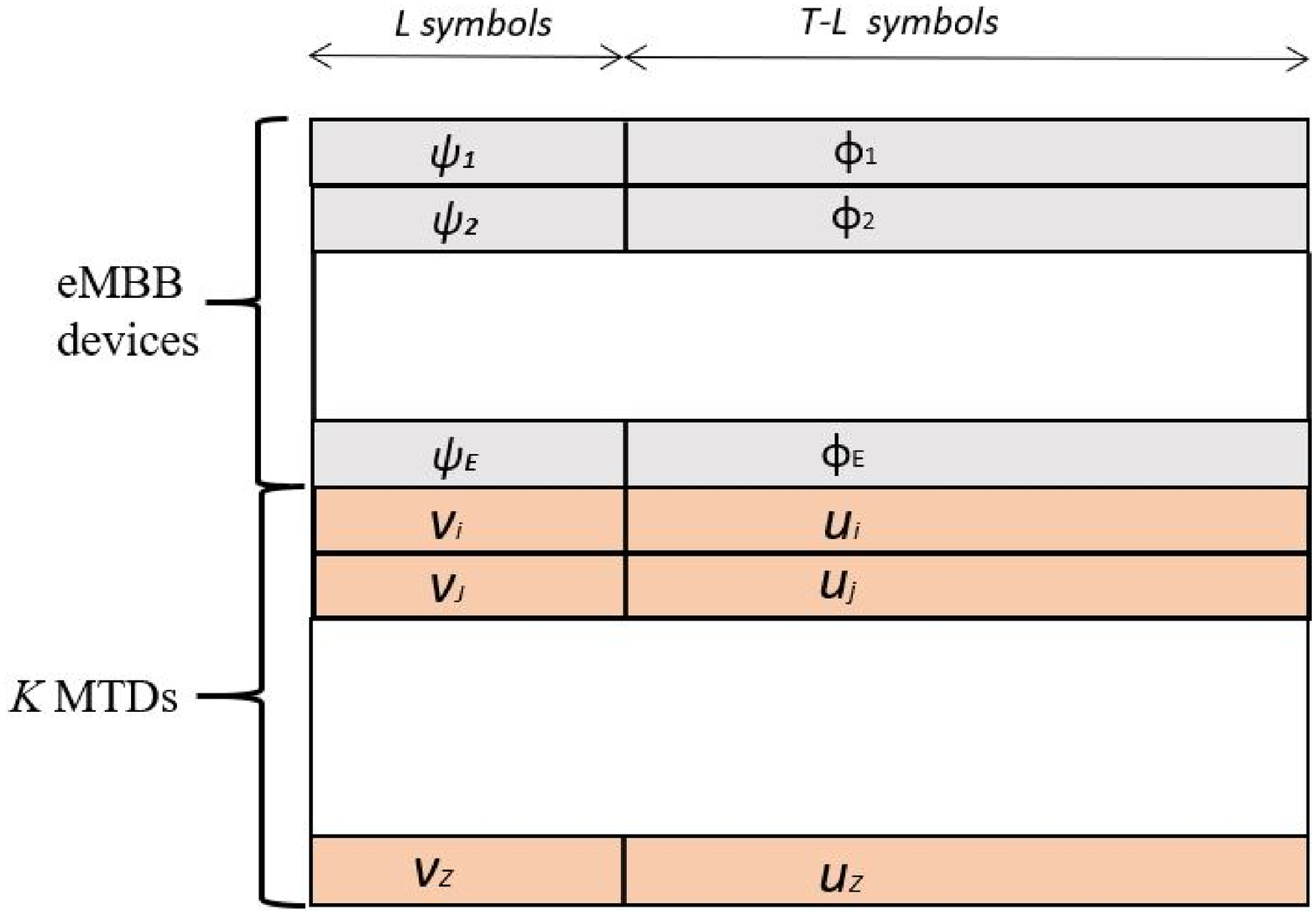}
  %\caption{Shared transmission block, where $i,j,\cdots,K \in \mathcal{K}$ and $\mathbf{u}_n\in \{\mathbf{u}_n^q\}_{q= 1,\cdots,Q}$}
   \label{figMTC2}
\end{subfigure}
%
%\vspace{-7mm}
\caption{ a) Network model (top), and b) shared transmission block (bottom) , where $i,j,\cdots,Z \in \mathcal{K}$ and $\mathbf{u}_n\in \{\mathbf{u}_n^q\}_{q= 1,\cdots,Q}$.}
\label{systemModel}%\vspace{-7mm}
\end{figure}
%\section{Receive Signal Processing}

\section{eMBB Channel Estimation and Decoding}
\label{detections}
To enable coherent decoding of eMBB devices, the CSI of each eMBB device is acquired using a pilot sequence of $L$ symbols, while $T-L$ symbols are used for encoding the actual intended messages, i.e., payload transmission. This two-phase strategy is chosen because it is efficient to perform CSI estimation followed by payload transmission in eMBB services \cite{senel2018grant}. The signal received at the BS is a combination of the signals from the $E$ eMBB devices and the active MTDs, with a working assumption that $E<L$. In the symbol slot $k\in \{1,2,\cdots,T\}$, the BS receives a signal $\Bar{\mathbf{y}}[k]\in\mathbb{C}^{M\times 1}$ of the form\begin{equation}
\Bar{\mathbf{y}}[k]\!=\! \sum_{e=1}^{E} \sqrt{\rho_e^{\mathrm{UL}}} s_{e}[k]\mathbf{h}_e \!+\! \sum_{n=1}^{N}\alpha_n s_{n}[k]\mathbf{g}_{n}\!+\! \mathbf{n}[k],
    \label{equatio1}
\end{equation}{\noindent where $\mathbf{g}_{n} = \sqrt{p_n^{\mathrm{UL}}}\tilde{\mathbf{g}}_n$, $\alpha_n\in\{0,1\}$ is the activity indicator for the $n$-th MTD, $\mathbf{n}[k]\sim \mathcal{CN}(\mathbf{0},\sigma^2\mathbf{I}) \in \mathbb{C}^{M \times 1}$ is the additive white Gaussian noise (AWGN) at the receiver antennas, $s_n[k]$ is the symbol transmitted by the $n$-th MTD in the $k$-th slot, with $\mathbb{E}\{(s_n[k])^2\} = \frac{1}{T}$, and $s_e[k]$ is the symbol transmitted  by the $e$-th eMBB device in the $k$-th slot, with $\mathbb{E}\{(s_e[k])^2\} = \frac{1}{T}$}. The transmitted symbols by the $e$-th eMBB device are split into two parts as
 \begin{equation}
   s_{e}[k]\!=\!\begin{cases} 
      \psi_{e}[k],~  \text{$k=\{1,2,\cdots,L\}$ } \\
      \phi_{e}[k],~  \text{$k=\{L+1,L+2,\cdots,T\}$}
   \end{cases}, \forall e \in \mathcal{E},
 \end{equation}where $ \bm{\psi}_{e} \in \mathbb{C}^{L \times 1}$ and $\bm{\phi}_{e} \in \mathbb{C}^{(T-L) \times 1}$ are vectors containing the pilot and payload data for the $e$-th eMBB device, respectively (refer to Fig. \ref{systemModel}(b)). At this stage, it is important to mention that for symbol slots ${k\in\{1,2,\cdots,L\}}$, \{$\bm{\psi}_{e}\}$ are mutually orthogonal, i.e., $\bm{\psi}_{i}^{T}\bm{\psi}_{j} = 0$, $i,j\in \mathcal{E}$,  with $i\ne j$, and also designed to be orthogonal to  $[s_n[1],s_n[2],\cdots,s_n[L]]\tran$, $\forall$ ${n\in\mathcal{N}}$. The details of this joint pilot and signature sequence design are provided in Section~\ref{sequence}. The orthogonality in slots ${k\in\{1,2,\cdots,L\}}$ is the key feature utilized to accurately acquire CSI estimates of the eMBB devices, as will be discussed next.  

%%%%%%%%%%%%%%%%%%
\subsection{Channel Estimation}
\label{csiEmbbF}
At the symbol slots $k=\{1,2,\cdots,L\}$, the $L$ received symbols at the BS are %given by the matrix
\begin{equation}
\Bar{\mathbf{Y}}_p=[\Bar{\mathbf{y}}[1],\Bar{\mathbf{y}}[2],\cdots,\Bar{\mathbf{y}}[L]]\tran
\label{pilotTrain}, 
\end{equation}
which are subsequently used in the correlation process to compute the received pilot signal from each eMBB device, $\mathbf{y}_e\in \mathbb{C}^{M \times 1}$, as 
\begin{equation}
    {\mathbf{y}_e =\Bar{\mathbf{Y}}_p\tran\bm{\psi}_{e}^{*}} = \mathbf{h}_{e}\bm{\psi}_{e}\tran\bm{\psi}_{e}^{*} +   \mathbf{N}_p\bm{\psi}_e^{*}, \qquad \forall e\in\mathcal{E}, 
    \label{equationCorelation}
\end{equation}
where $\mathbf{N}_p = \left[\mathbf{n}[1], \mathbf{n}[2],\cdots,\mathbf{n}[L]    \right] \in \mathbb{C}^{M \times L}$. 
Note that due to the mutual orthogonality of $\{\bm{\psi}_{e}\}$ and orthogonality to the first $L$ symbols transmitted by the MTDs, the processed signal $\mathbf{y}_e$ is free from interference of other eMBB devices and the MTDs. This in turn enables accurate estimation of the CSI of the $e$-th eMBB device, thus promoting coexistence of eMBB and MTDs. 

To obtain a CSI estimate of the $e$-th eMBB device, the BS uses the MMSE estimator % defined by
\cite{ozdogan2019massive}
\begin{equation}
 \hat{\mathbf{h}}_e = \sqrt{\rho_{e}^{\mathrm{UL}}}\beta_{e}\bm{\Pi}_{e}\mathbf{y}_e , 
\end{equation}
where $\bm{\Pi}_{e} = (\sigma^2 + \beta_e)^{-1}\mathbf{I} \in \mathbb{C}^{M \times M }$ is the covariance matrix of $\mathbf{y}_e$. It then follows that the MMSE estimate, $\hat{\mathbf{h}}_e$, and the estimation error $\tilde{\mathbf{h}}_e = \hat{\mathbf{h}}_e - \mathbf{h}_e$ are distributed as \cite{ozdogan2019massive}\begin{align}
  \hat{\mathbf{h}}_e&\sim \mathcal{CN}(\mathbf{0},\beta_{e}\mathbf{I}-   \mathbf{\Xi}_{e}),\quad 
  \tilde{\mathbf{h}}_e \sim \mathcal{CN}(\mathbf{0},   \mathbf{\Xi}_{e}), 
  \label{resnoise}
  \nonumber
\end{align}
respectively, where $\mathbf{\Xi}_{e} \in \mathbb{C}^{M\times M}$ is the covariance matrix of the estimation error, defined as
\begin{equation}
    \mathbf{\Xi}_{e} = \beta_{e}\mathbf{I}-L\rho_{e}^{\mathrm{UL}}\beta_{e}^2\bm{\Pi}_{e}.
\end{equation}

% We assume that the BS uses $\hat{\mathbf{h}}_e$ to decode the eMBB data using \textcolor{blue}{the MMSE combiner} \cite{albreem2021overview}. 
\subsection{Data Decoding} %  of the eMBB Device
\label{embbsDatadecoding}
 We assume that the BS uses the MMSE combiner  \cite{albreem2021overview} to decode the signals from the eMBB devices. In general, the BS' ability to accurately decode the eMBB signals is subject to the accuracy of the channel estimate $\mathbf{\hat{h}}_{e}$. This impact can be captured by the signal-to-interference plus noise ratio (SINR), which for each  eMBB device is given by \eqref{equSNIR} at the top of the next page, where $\bm{\omega}_e  = \bigl(\sigma^2\mathbf{I} + \sum\nolimits_{j\in\mathcal{E}\neq e}\mathbf{\hat{h}}_{j}\mathbf{\hat{h}}_{j}\herm\bigr)^{-1}\mathbf{\hat{h}}_{e}$ is the receive combiner. The value of $\bm{\omega}_e$ depends on $\hat{\mathbf{h}}_e$ and therefore introduces the estimation error into the SINR.
 \begin{figure*}[t!]
 \begin{align}
     \mathrm{\Gamma}_{e} =  \frac{ \sum\limits_{k = L +1}^{T}\sqrt{\rho_e^{\mathrm{UL}}}|\bm{\omega}_e\herm\hat{\mathbf{h}}_e s_e[k]|^2}{ \sum\limits_{k = L +1}^{T} \left(\sum\limits_{\forall i\in \mathcal{K}}\alpha_i|\bm{\omega}_e\herm\mathbf{g}_{i}s_i[k]|^2 + \sum\limits_{\forall j\in \mathcal{E} \neq e}\sqrt{\rho_j^{\mathrm{UL}}}|\bm{\omega}_e\herm\mathbf{h}_{j}s_{j}[k]|^2  + \sqrt{\rho_e^{\mathrm{UL}}}|\bm{\omega}_e\herm\tilde{\mathbf{h}}_es_e[k]|^2 + |\bm{\omega}_e\herm\mathbf{n}[k]|^2 \right)}\label{equSNIR} \\
     \bottomrule\nonumber
     \end{align}
 \end{figure*}
     
 The expression \eqref{equSNIR} is defined differently from works such as \cite{senel2018grant} because we assume that the symbols can be correlated, e.g., MTDs can have similar messages, as will be seen in Section~\ref{sequence}. Nevertheless, $\mathrm{\Gamma}_{e}$, can be interpreted as the measure of how much the signal from the $e$-th eMBB device is buried in the network. As a result, the probability of decoding errors tends to be lower for devices with relatively high values of $\mathrm{\Gamma}_{e}$ as compared to those with lower values. To facilitate non-coherent decoding of the MTDs, the BS removes signal contribution of devices that were correctly decoded from  $\Bar{\mathbf{y}}$ in (\ref{equatio1}) using SIC. As a result, the coexistence of the eMBB devices and MTDs, is affected by how accurate the SIC is performed. 
\section{Joint Activity Detection and Decoding of MTDs}
\label{JADD}
\subsection{Non-coherent Decoding of MTDs}
Non-coherent signal recovery promises to reduce the access latency caused by the need to acquire explicit CSI before data decoding \cite{senel2018grant,ngo2020multi}. This is specifically appealing for short packet communications such as in MTC. In order to convey data non-coherently from MTDs, each of the MTDs is pre-allocated $Q$ different messages. Under this scenario, each active MTD can strictly  transmit one selected message in a given coherence interval. From CS-MUD perspective, recovery of these messages can be achieved by joint activity and data decoding (JADD).  

It is important to highlight that due to the heterogeneity of the network, the BS recovers the messages from the MTDs after $T$ symbols have collectively been used for CSI estimation, decoding, and SIC of correctly decoded eMBB signals. Ideally, after processing the eMBB data, the remaining signal should be  coming purely from the MTDs. However, this will rarely be the case, because CSI estimation is normally imperfect and thus leads to imperfect SIC. Moreover, the number of MTDs can be massive while $T$ is limited. Considering this, it is apparent that the BS will use very few symbols to recover a large number of sequences because $T\ll N$. Mathematically, this poses the JADD process as an underdetermined system\footnote{An underdetermined system of linear equation has more unknowns than equations \cite{wang2010unique}.}, which motivates us to model the JADD as a compressed sensing problem. Next, we present the sparsity pattern of data from MTDs, which lays foundation to the JADD problem.

The central idea of the JADD is to unambiguously identify each device's transmitted data. Let the collection of each MTD's sequences be defined as  \begin{equation}
    \mathbf{S}_n = [\mathbf{s}_{n}^1,\mathbf{s}_n^2,\cdots,s_{n}^Q], 
\end{equation}
where $\mathbf{s}_{n}^q \in \mathbb{C}^{T \times 1}$
corresponds to the $q$-th signature sequence. In light of this, the JADD problem should be posed such that the BS can identify each of the message sequences sent by active MTDs. To take this into consideration, we define the sequence transmission indicator associated with the $q$-th sequence of the $n$-th MTDs by    
\begin{equation}
  \alpha_n^q \!=\!
  \begin{cases} 
      \!1, \text{if $q-$th sequence of MTD $n\!\in\!\mathcal{N}$ is transmitted}, \\
      \!0,\text{otherwise,}\\
   \end{cases} 
   \label{truedetec}
\end{equation}
and let ${\bm{\alpha} =[\alpha_1^1,\cdots,\alpha_1^Q,\cdots,\alpha_N^q,\cdots,\alpha_N^Q]\tran} \in \mathbb\{0,1\}^{NQ\times 1}$ be a complete collection of sequence transmission indicators for all $N$ devices. This variable plays a crucial role in the JADD problem, as we will see next. 
%%% This paragraphs discusses JADD problem in detail to tie the estimation problem to the sequences.

From above, each MTD has a dictionary of messages. However, since each active MTD can transmit one message sequence in a coherence interval, each of the active MTDs has one non-zero $\alpha_n^q$. This introduces the following relation between the device activity indicator $\alpha_n$ (see \eqref{equatio1}) and the sequence transmission indicator $\alpha_n^q$ in \eqref{truedetec}: if device $n$ is active (i.e., ${\alpha_n=1}$) and it transmits its $q$-th message (i.e., ${\alpha_n^q =1}$), then ${\alpha_n =\alpha_n^q = 1}$. This reveals that it is impossible to identify a transmitted sequence by solely using $\alpha_n$. In view of this, the JADD problem needs to be posed to find the non-zero $\alpha_n^q$'s, thus jointly identifying an active device $n$ and its $q$-th transmitted message sequence. In summary, the BS performs JADD by estimating $\bm{\alpha}$.

%To \blue{estimate $\bm{\alpha}$}, we formulate JADD using the signal at the BS's disposal after SIC of the eMBB data as 
In order to perform JADD via estimation of $\bm{\alpha}$, we first formulate the signal at the BS's disposal after SIC of the eMBB data as 
\begin{equation}
    \mathbf{Y} =\sum_{n=1}^{N} \sum_{q=1}^{Q}\alpha_{n}^{q}\mathbf{s}_n^{q}\mathbf{g}_{n}\tran +\mathbf{W}, 
    \label{eq223}
\end{equation}
where $\mathbf{W}\in \mathbb{C}^{T\times M}$ comprises both the receiver noise and the resultant residual interference from imperfect SIC (due to errors in the CSI estimation of the eMBB devices). The $m-$th column of $\mathbf{W}$ is denoted by ${\mathbf{w}_{m} = [(\bm{w}_{p})\tran, (\bm{w}_{d})\tran]\tran}$, where $\bm{w}_{p}\sim \mathcal{CN}(\mathbf{0},\sigma^2\mathbf{I} + \sum_{e =1}^{E}\mathbf{\Xi}_{e}) \in \mathbb{C}^{L\times 1}$ is the combination of the AWGN and the error vector during eMBB training and $\bm{w}_{d}\in \mathbb{C}^{(T-L) \times 1}$ models the combination of the AWGN and the possible error vector due to eMBB decoding followed by imperfect SIC. Note that the structure of $\bm{w}_d$ depends on the accuracy of SIC, as discussed in Section \ref{embbsDatadecoding}. It is therefore important to note that even though there will be varying degrees of accuracy of SIC, there are two special cases: i) when SIC is perfect, $\bm{w}_{d}$ reduces to AWGN and ii) when SIC is not performed, $\bm{w}_{d}$ reduces to the deterministic composite signal of eMBB devices, i.e.,  $\sum_{e=1}^{E}\sqrt{\rho_e^{\mathrm{UL}}}\bm{\phi}_e$. 
%\textcolor{blue}{However, it is also crucial to highlight that in between these two extremes there is infinite number of values that can be taken by $\bm{w}_{d}$ depending on the accuracy of SIC.}

From the preceding discussion, the BS processes $\mathbf{Y}$ with the intention to recover $\bm{\alpha}$. To reveal the sparsity pattern in $\bm{\alpha}$, we rewrite the signal \eqref{eq223} in its matrix-equivalent form as
\begin{align}\label{mtc}
 \mathbf{Y} &= \sum_{n=1}^{N}[\mathbf{s}_{n}^{1},\mathbf{s}_{n}^{2},\cdots,\mathbf{s}_{n}^{Q}]\underbrace{\begin{bmatrix}
   \alpha_n^{1} & & \\
    & \ddots & \\
    & & \alpha_n^{Q}
  \end{bmatrix}}_{\mathbf{D}_n}\underbrace{\begin{bmatrix}
   \mathbf{g}_{n}\tran \\
     \mathbf{g}_{n}\tran \\
     \vdots\\ 
     \mathbf{g}_{n}\tran
  \end{bmatrix}}_{\mathbf{G}_n}
  +\mathbf{W} \nonumber \\
  &= \sum_{n=1}^{N}\mathbf{S}_{n}\mathbf{D}_{n}\mathbf{G}_{n}+\mathbf{W} =\mathbf{S}\mathbf{D}\mathbf{G} + \mathbf{W},
\end{align}
where ${\mathbf{S} =[\mathbf{S}_{1},\mathbf{S}_{2},\cdots,\mathbf{S}_{N}]\in\mathbb{C}^{T\times NQ}}$ is a collection of all the message sequences allocated to the MTDs,  ${\mathbf{D}=\mathrm{diag}(\mathbf{D}_1,\mathbf{D}_2,\cdots, \mathbf{D}_{N})\in \mathbb{R}^{NQ\times NQ}}$ is the (sequence) activity indicator matrix with ${\mathbf{D}_n=\mathrm{diag}\{\alpha_{n}^1,\alpha_{n}^2,\cdots,\alpha_{n}^Q\}\in\{0,1\}^{Q\times Q}}$, and ${\mathbf{G} = [\mathbf{G}_{1}\tran, \mathbf{G}_{2}\tran,\cdots,\mathbf{G}_{n}\tran]\tran \in \mathbb{C}^{NQ\times M}}$, where $\mathbf{G}_n$ is a matrix comprising repeated effective channels of the $n$-th MTD. The received signal is %therefore given by 
\begin{equation}
    \mathbf{Y} = \mathbf{S}\mathbf{X}+ \mathbf{W},
    \label{sparsityVec}
\end{equation}
where $\mathbf{X}=\mathbf{D}\mathbf{G}\in \mathbb{C}^{NQ\times M}$ is a row-sparse matrix. 

Explicitly, the sparse rows of $\mathbf{X}$ can be represented by 
\begin{equation}
\mathbf{X}\!=\!
\begin{bmatrix}
\mathbf{X}_{1}\\
\mathbf{X}_{2}\\
\vdots\\
\mathbf{X}_{N}
\end{bmatrix}
\!=\!
\begin{bmatrix}
\mathbf{x}_{11}\tran\\
\mathbf{x}_{12}\tran\\
\vdots\\
\mathbf{x}_{1Q}\tran\\
\mathbf{x}_{21}\tran\\
\mathbf{x}_{22}\tran\\
\vdots\\
\mathbf{x}_{2Q}\tran\\
\vdots\\
\mathbf{x}_{nq}\tran\\
\vdots\\
\mathbf{x}_{NQ}\tran
\end{bmatrix}
\!=\!
\begin{bmatrix}
\alpha_1^1[{g}_{11},{g}_{12},\ldots,{g}_{1M}]\\
\alpha_1^2[{g}_{11},{g}_{12},\ldots,{g}_{1M}]\\
\vdots\\
\alpha_1^Q[{g}_{11},{g}_{12},\ldots,{g}_{1M}]\\
\alpha_2^1[{g}_{21},{g}_{22},\ldots,{g}_{2M}]\\
\alpha_2^2[{g}_{21},{g}_{22},\ldots,{g}_{2M}]\\
\vdots\\
\alpha_2^Q[{g}_{21},{g}_{22},\ldots,{g}_{2M}]\\
\vdots\\\alpha_n^q[{g}_{n1},{g}_{n2},\ldots,{g}_{nM}]\\
\vdots\\
\alpha_N^Q[{g}_{N1},{g}_{N2},\ldots,{g}_{NM}]\\
\end{bmatrix},
\label{newSparse}
\end{equation}{\noindent where ${\mathbf{X}_{n} = \mathbf{D}_n\mathbf{G}_n \in \mathbb{C}^{Q \times M}}$ is the $n$-th sub-matrix of $\mathbf{X}$ and ${\mathbf{x}_{nq}\tran = \alpha_n^q\mathbf{g}_{n}\tran}$ is the effective channel of the $n$-th MTD when transmitting its $q$-th sequence, thus, corresponding to row number $(n-1)Q+q$ of $\mathbf{X}$. Recall that the row sparsity of $\mathbf{X}$ has a special structure: each $\mathbf{X}_{n}$ can have at most one non-zero row.}
Note that, with the structure in (\ref{newSparse}), the non-zero entries of $\mathbf{X}$ occur with probability $\xi = \frac{\epsilon}{Q}$, even though the activation probability of MTDs is $\epsilon$. Alternatively, by using this row sparsity, each sequence can be interpreted as a fictitious device.    

From (\ref{sparsityVec}), $\mathbf{Y}\in \mathbb{C}^{T \times M}$ is a noisy compressed measurement of $\mathbf{X}$. It then follows that the JADD task is to recover $\mathbf{X}$ from $\mathbf{Y}$ with the knowledge of $\mathbf{S}$ by using some sparsity-promoting operation, $f(\cdot)$, resulting in an estimate of $\mathbf{X}$ as $\hat{\mathbf{X}}=f(\mathbf{Y};\mathbf{S})$. From this view point, the efficacy of $f(\cdot)$ is inherently improved by optimizing $\mathbf{S}$. To this end, we propose an algorithm for generating message sequences $\mathbf{S}$ for MTDs in Section~\ref{sequence}, while the formal JADD problem and its algorithmic solutions are presented in Section~\ref{SSRALG}.

%%%%%%%%%%%%%%%%%%%%%%%%%
\subsection{Joint Generation of eMBB Pilots and MTD Sequences}
\label{sequence}
Like motivated in Section~\ref{csiEmbbF}, successful coexistence of eMBB and MTC traffic requires that each pilot sequence $\bm{\psi}_e$ be orthogonal to the data of the MTDs in the first $L$ symbol slots. Consistent with this requirement, we propose a joint generation of $\mathbf{S}$ and $\{\bm{\psi}_e\}$.  

First, we handle the assignment of the pilot sequences to the $E$ eMBB devices. We generate $L$ orthogonal sequences $\left[\mathbf{c}_{1},\mathbf{c} _{2},\cdots,\mathbf{c}_{L}\right] \in \mathbb{C}^{L \times L}$ using a complex-valued Hadamard matrix\footnote{Such sequences are basically quadrature amplitude modulation (QAM) symbols.}, whose columns and rows are mutually orthogonal, i.e., ${\mathbf{c}_{i}\herm\mathbf{c}_{j} = 0, \forall i\neq j}$. Then, without loss of generality, we let $ \bm{\psi}_{1} = \mathbf{c} _{1},\bm{\psi}_{2} = \mathbf{c} _{2}, \cdots,\bm{\psi}_{E} = \mathbf{c} _{E}$. 

Once $\{\bm{\psi}_{e}\}$ are assigned to the $E$ eMBB devices, then there are ${L-E}$ orthogonal sequences remaining to be allocated to the MTDs for the first $L$ symbol slots. To facilitate the presentation, we decompose each message sequence of the $n$-th MTD as $\mathbf{s}_n^q=\left[(\mathbf{v}_n)\tran,(\mathbf{u}_{n}^{q})\tran \right]\tran$, ${q=1,\cdots,Q}$, where ${\mathbf{v}_n \in \mathbb{C}^{L\times 1}}$ and ${\mathbf{u}_{n}^{q}\in \mathbb{C}^{(T-L) \times 1}}$ are the message (sub)sequences for slots $\{1,\cdots,L\}$ and ${\{L+1,\cdots,T\}}$, respectively. Note that the message sequences of the $n$-th MTD, $\mathbf{s}_n^q$, ${q=1,\cdots,Q}$, share the same ``header'' $\mathbf{v}_n$, while their second parts $\mathbf{u}_{n}^{q}$ are unique. Next, we address the design of each $\mathbf{v}_n$, ${n\in\mathcal{N}}$. 

Given that ${L-E\ll N}$, it is not feasible to allocate orthogonal sequences to each MTD to create $\mathbf{v}_n$, ${n\in\mathcal{N}}$. In addition, the requirement that $\{\mathbf{v}_n\}$ has to be orthogonal to $\{\bm{\psi}_e\}$ makes it necessary to reuse the remaining $L-E$ orthogonal pilot sequences for multiple MTDs. However, this poses a risk of possible high correlations between the pilot sequences $\{\mathbf{v}_n\}$. Such highly correlated pilot sequences can lead to ambiguities during identification of the MTDs by means of mutual interference. Evidently, these ambiguities result in degradation in the performance of the receiver. To avert this challenge, we devise a pilot design strategy that can mitigate the mutual interference between the MTDs by maintaining high incoherence between each $\mathbf{v}_n$. We do this as follows. 
Let ${\mathcal{B}=\{\mathbf{c}_{E +1},\cdots,\mathbf{c}_{L}}\}$ be the set of the ${L-E}$ remaining orthogonal sequences after assigning  $\{\bm{\psi}_e\}$ to the eMBB devices. The sequence $\mathbf{v}_n$ of the $n$-th MTD is formed by randomly selecting a subset of orthogonal sequences in $\mathcal{B}$ which are then linearly combined using random weights\footnote{Power normalization of the entire message sequence $\mathbf{s}_n^q$ is done later.}. More precisely, let $\pi_n=\{\pi_n(1),\cdots,\pi_n(z)\}$ be a collection of indices associated with the random sequence selection, where ${\pi_n(l)\in\{E+1,\ldots,L\}.}$ Here, ${z < L}$ is a parameter that adjusts how many sequences each MTD can combine. Note that as $z$ increases, more orthogonal columns are combined and as such, more highly correlated pilot sequences are generated. We address this by minimizing the probability that two devices have the same $\mathbf{v}_n$, hence ensuring that each MTD can be identified without ambiguity. To achieve this, we fix a reasonable maximum collision probability \cite{senel2018grant} and find the minimum value of $z$ that attains this probability. Hence, we solve an integer problem 
\begin{equation}
  \begin{array}{ll}
\underset{z\in\{1,2,3,\ldots\}}{\operatorname*{minimize}} \quad & z \\
\textrm{subject to} \quad & \frac{1}{\binom{L-E}{z}}\leq \chi,
\end{array}
\label{zOpt}
\end{equation}{\noindent where $\chi$ is the target collision probability\footnote{It is important to note that the value of $\chi$ is high for small $L$, and thus impractical to set similar values of $\chi$ for all values of $L$. For example, for $L = 8$, the lowest achievable collision probability is $\chi = 0.1$. In this case, $z$ is set to any value below $L-E$. For $L\geq 32$, a common $\chi$ can be used.}.} 
% \textcolor{blue}{We highlight that to solve \eqref{zOpt}, a value of $\chi$ is  a design parameter which can be set similar or close to values reported in \cite{senel2018grant}.} 

% %%
% \textcolor{red}{  To avert this challenge, our work sets out to find a pilot design strategy that can mitigate complete degradation. This is done by randomly combining $z$ orthogonal columns from the $E-L$ remaining ones. Note that as $z$ increases, more orthogonal columns are combined and as such more correlated pilot sequences are generated. We address this by minimizing the probability that two devices have the same $\mathbf{v}_n$ hence ensuring that each MTD can be identified without ambiguity. To achieve this, we fix a reasonable collision probability \cite{senel2018grant} and find the minimum value of $z$ that meets this probability.}
%%

\begin{figure}[t]
    \centering
    \includegraphics[width=0.45\textwidth]{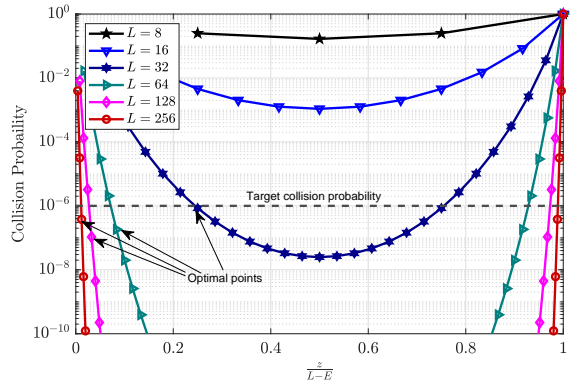}
\caption{Collision probability as a function of $z$ for different pilot lengths.}
    \label{optimalZ}
\end{figure}

\begin{figure}[t]
    \centering
    \includegraphics[width=0.45\textwidth]{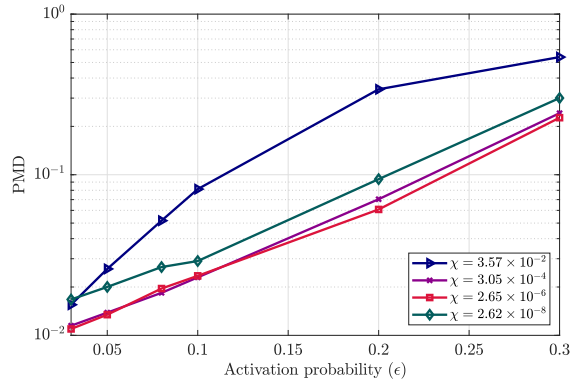}
\caption{PMD as a function of $\epsilon$, for different values of $\chi$ using EM-SBL, $\text{PFA} = 10^{-3}$, $L = 32$.}
    \label{optimalZMissedDetect}
\end{figure}

% \textcolor{blue}{Observe that different values of $z$ yield different values of collision probability for a given plot of $L$}.
The problem \eqref{zOpt} does not have a closed-form solution. We therefore exhaustively search over possible values of $z$, and pick the minimum value of $z$ that satisfy the collision probability constraint. Note that the search is done simply over the positive integers and is thus of low complexity.  Fig.~\ref{optimalZ} illustrates the collision probability as a function of $z$, for $E = 4$. The values of $z$ that solve \eqref{zOpt} for the targeted collision probability are marked. Note that for fixed $L$, different values of $z$ yield different collision probabilities ($\chi$).  In Fig.~\ref{optimalZMissedDetect} we quantify the impact of $\chi$ on the receiver's ability to detect the MTDs. This is done by evaluating the PMD as a function of $\chi$ and the average activation probability ($\epsilon$), for $\text{PFA} = 10^{-3}$ using SBL\footnote{The details of SBL will be presented later in Section~\ref{SSRALG}.}. In the figure, the lowest PMD is reported for ${\chi = 2.65\times 10^{-6}}$, while the worst is for $\chi = 3.75\times 10^{-2}$. We also note that, even though ${\chi = 2.62\times 10^{-8}}$ is the lowest collision probability for $L = 32$, its performance is worse than that of $\chi = 2.65\times 10^{-6}$. This is so because, to achieve ${\chi = 2.62\times 10^{-8}}$ the value of $z$ has to be set to $\frac{L-E}{2}$, hence resulting in more correlated $\{\mathbf{v}_n\}$. 

Finally, the $n$-th MTD is assigned $\mathbf{v}_n$ by combining a collection of sequences\\ ${\mathcal{B}_n = \{\mathbf{c} _{\pi_n(1)},\cdots,\mathbf{c}_{\pi_n(z)}\}\subseteq\mathcal{B}}$ as
\begin{equation}\label{eq:vn}
\mathbf{v}_n =\sum_{l=1}^{z}\vartheta_{n,l}\mathbf{c}_{\pi_n(l)},
\end{equation}
where $\vartheta_{n,l}$, ${l=1,\ldots,z}$, are non-negative combining weights. To conclude, by constructing $\mathbf{v}_n$ as given in \eqref{eq:vn}, we ensure the required orthogonality by having $\mathbf{v}_n\tran\bm{\psi}_e = 0, \forall n\in \mathcal{N}, \forall e \in \mathcal{E}$. Next, we present {the} optimization of the second part of the sub-sequences of each MTD. 

To further enhance the JADD of the MTDs, sub-sequences (messages) $\mathbf{u}_n^q$, $q=1,\dots,Q$, of a device $n$ are chosen such that they maintain minimal cross-correlation among one another. Our proposed sequence optimization is presented in \textbf{Algorithm}~1. In Line 1, the algorithm is initialized by a matrix $\tilde{\mathbf{U}}_{n}\in\mathbb{C}^{\kappa^{T-L}\times(T-L)}$, which contains all possible candidate sequences in its rows, generated from the modulation alphabet of order $\kappa$. In Line 2, the cross-correlation matrix $\mathbf{\Theta}$ is computed, which captures the amount of correlation between any pairs of candidate sequences in $\tilde{\mathbf{U}}_{n}$. Subsequently, the iterative procedure in Lines 4--6 utilizes $\mathbf{\Theta}$ to assign unique sequences (i.e., rows of $\tilde{\mathbf{U}}_{n}$) that yield the minimal cross-correlation. Line 4 gives the $i^*$-th and $j$-th rows that are least correlated. In Line 5, the $q$-th sub-sequence of MTD $n$ is assigned based on the index $i^*$, i.e., $\mathbf{u}_n^q=\tilde{\mathbf{U}}_{n,i^*}$.
%Line 6 prevents the reuse of $i^*$-th sequence of $\tilde{\mathbf{U}}_{n}$.
Line 6 prevents reuse of the $i^*$-th sequence of $\tilde{\mathbf{U}}_{n}$. 
The complexity of the proposed algorithm is ${\mathcal{O}\left(\left(\frac{(\kappa^{T-L})(\kappa^{T-L}-1)  }{2}\right)^2 + (T-L)\kappa^{T-L}\right)}$. Evidently, this is a high complexity algorithm. However, since MTDs generally have constant messages, the algorithm can be run once and only updated when a new device is added to the network. 

\begin{algorithm}[t]
\textbf{Input:} A matrix of all possible messages  $\tilde{\mathbf{U}}_{n}=[\tilde{\mathbf{u}}_n^1,\tilde{\mathbf{u}}_n^2,\cdots,\tilde{\mathbf{u}}_n^{^{\kappa^{T-L}}}]\tran\in\mathbb{C}^{\kappa^{T-L}\times(T-L)}$
\\\
Compute $\bm{\Theta}=|\tilde{\mathbf{U}}_n\tran\tilde{\mathbf{U}}_n|$.
\\\
\textbf{For $q=1,\ldots,Q$:}
\\\
$\{i^*,j\}=\underset{\substack{i,j\in\{1,2,\ldots,\kappa^{T-L}\},~i>j}}{\mathrm{arg~min}}~\bm{\Theta}_{i,j}$
\\\
$\mathbf{u}_n^q=\tilde{\mathbf{U}}_{n,i^*}$
\\\
Set $\bm{\Theta}_{i^*,j}=\infty$ for all $j=\{1,2,\ldots,\kappa^{T-L}\}$.
\caption{Generation of messages $\mathbf{u}_n^q$, $q=1,\dots,Q$ for each MTD $n$}
\textbf{End for}
\label{messagesGen}
\end{algorithm}

%%%%%%%%%%%%%%%%%%%%%%%%%%%%%%%%%%%%%%%%
\section{Sparse Signal Recovery Algorithms for JADD}\label{SSRALG} %ADD solution via sparse recovery
%\section{Sparse Row $\ell_{2,1}$ minimization }\label{probformulate}
To perform JADD after employing {the} SIC of the decoded eMBB signals, it should be noted that the effective channel matrix $\mathbf{X}$ in \eqref{newSparse} is strictly (row-)sparse and can be recovered using SSR techniques for a multiple measurement vector (MMV) problem. In our problem, we note the following special structures of $\mathbf{X}$:  i) the rows of $\mathbf{X}$ are sparse as shown in (\ref{newSparse}), and ii)  each $\mathbf{D}_n$ has at most one non-zero element in its diagonal, hence imposing a restriction that at most one row of $\mathbf{X}_n$ is non-zero. Taking this structure into consideration, the canonical form for the JADD problem can be formulated as \begin{subequations} \begin{alignat}{2}
       \operatorname*{minimize}_{\mathbf{X}}& \quad  \frac{1}{2}\norm{\mathbf{Y}-\mathbf{S}\mathbf{X}}_{F}^2 + \mu \norm{\mathbf{X}}_{2,0} \label{x_0} \\
      \text{subject to} & \quad  \mathrm{Rank}(\mathbf{D}_{n})\leq 1,\forall n \in \mathcal{N} \label{Rank}, 
  \end{alignat}
   \label{l_2,0}
\end{subequations}where ${\mathbf{X}_{2,0} = \norm{\norm{\mathbf{x}_{11}}_2,\cdots,\norm{\mathbf{x}_{NQ}}_2}_0} $ enforces the sparsity along the rows of $\mathbf{X}$ and $\mu>0$ is a regularization parameter. However, we note that the problem \eqref{l_2,0} is non-convex and thus hard to solve. The non-convexity arises due to the fact that i) the objective function \eqref{x_0} gives rise to an  $\ell_0$-norm minimization problem, which is generally NP-hard \cite{zheng2021jointMMV,eldar2012compressed}, and ii) the rank constraint\footnote{In this work, we do not pursue for a tractable (convex) penalty function to incorporate the rank constraint into the optimization process. We instead use a two-step simplification which incorporates the rank constraint by hard-thresholding.} \eqref{Rank} is non-convex.

%%%%%%%%%%%%%%%

%%%%%%%%%%%%%%

To provide solution to problem \eqref{l_2,0}, the JADD problem is solved in two steps. First, we consider tractable solutions to the problem \eqref{x_0} by neglecting constraint \eqref{Rank}. Thereafter, an estimate of $\ell_2$-norms of each row of $\hat{\mathbf{X}}$ is computed at the BS as 
\begin{equation}
    \Bar{\mathbf{x}} = [\norm{\hat{\mathbf{x}}_{11}}_{2},\norm{\hat{\mathbf{x}}_{12}}_{2},\cdots, \norm{ \hat{\mathbf{x}}_{nq}}_{2},..,\norm{\hat{\mathbf{x}}_{NQ}}_{2}]\tran
    \label{l21sdp}.
\end{equation}
Let $\Bar{\mathbf{x}}_{n} = [\norm{\hat{\mathbf{x}}_{n1}}_{2},\cdots, \norm{\hat{\mathbf{x}}_{nq}}_{2},..,\norm{\hat{\mathbf{x}}_{nQ}}_{2}]\tran \in \mathbb{R}^{Q \times 1}$ be the norms of all the $Q$ rows corresponding to the $n-$th MTD. The constraint \eqref{Rank} is incorporated by thresholding\footnote{We select the threshold based on the ROC. Precisely, we select a PFA from the ROC and use the threshold corresponding to this PFA value.} the maximum entry of $\Bar{\mathbf{x}}_{n}$ using $\zeta$. Hence,the estimated sequence transmission indicator for each $q$-th sequence is% estimated by 
\begin{equation}
  \hat{\alpha}_{n}^{q} = 
  \begin{cases} 
      1,~\text{if $ \mathrm{max}\{\Bar{\mathbf{x}}_n\}\geq \zeta$} \\
      0,~\text{otherwise}\\
   \end{cases}. 
   \label{maximumThreshold}
\end{equation}

Given a potentially massive number of MTDs and that each can be allocated more than one message sequence, the problem of solving for $\mathbf{X}$ is high-dimensional. As a remedy, we exploit computationally efficient iterative SSR procedures\footnote{In the present work, where our main focus is to lay foundation to the framework enabling coexistence of eMBB and MTC traffic, we do not tailor and optimise the SSR algorithms specifically to the underlying problem but merely exploit existing solutions. The design of more sophisticated SSR algorithms, which do not rely on the two-step procedure to neglect the rank constraint in the iterative optimization, is left for future work.} to solve the problem. The iterative procedures will perform the task of the operation $\hat{\mathbf{X}}=f(\mathbf{Y};\mathbf{S})$, thus solving (\ref{l_2,0}) approximately. In particular, the AMP algorithm \cite{chen2018sparse,wei2018joint,senel2018grant}, the ADMM based $\ell_{2,1}$-minimization algorithm \cite{boyd2011distributed,djelouat2021spatial}, EM-SBL \cite{wipf2004sparse}  and SOMP \cite{tropp2006algorithms} are discussed next. Moreover, since a typical 5GB cellular network is composed of a massive number of MTDs coexisting with eMBB devices, then  the computational complexity of the algorithms running at the BS is of critical importance as it influences power consumption and contributes to the carbon footprint. Table \ref{complexAnalysis} presents  a  summary of the complexity of each SSR algorithm.

%%%%%%%%%%%%%%%%%%%%%%%%%%%%%%%%%%%%%%%
\subsection{Solution via Approximate Message Passing (AMP)}
\label{solutionAMP}
The AMP utilised in this work is based on the Bayesian denoiser \cite{senel2018grant,chen2018sparse,wei2018joint}. From statistical point of view, AMP enforces sparsity by treating each row of $\mathbf{X}$ as a vector independently generated from the distribution
\begin{equation}
    p(\mathbf{x}_{nq}) = (1-\xi)\delta_{0}(\mathbf{x}_{nq})  + \xi\mathcal{CN}(\mathbf{x}_{nq};\mathbf{0},\gamma_n\mathbf{I}), 
    \label{priorDistribution}
\end{equation}
where $\delta_{0}(\mathbf{x}_{nq})$ is the Dirac delta function defined as
\begin{equation}
\delta_{0}(\mathbf{x}_{nq})=
\begin{cases}
\infty,&\text{if}~\mathbf{x}_{nq}=\mathbf{0}\\
0, &\text{otherwise}
\end{cases}
\end{equation}
and satisfying 
%\begin{equation} 
$\int_{\mathbf{x}_{nq}}\delta_{0}(\mathbf{x}_{nq})\mathrm{d}\mathbf{x}_{nq}=1$. 
%\end{equation}

%\blue{
%$\delta(\vec{x}_i)$ is the Dirac %delta function defined as
%\begin{equation}
%\delta(\vec{x}_i)=
%\begin{cases}
%\infty,~\text{if}\,\vec{x}_i=\vec{%0}\\
%0,~\text{otherwise},
%\end{cases}
%\end{equation}
%and satisfying 
%\begin{equation} 
%\int_{\vec{x}_i}\delta(\vec{x}_i)\%mathrm{d}\vec{x}_i=1. 
%%}

From (\ref{priorDistribution}),  $\mathbf{x}_{nq} = \mathbf{0}$ with probability $1-\xi$ and $\mathbf{x}_{nq} \sim \mathcal{CN}(0,\gamma_n\mathbf{I})$ with probability $\xi$. Based on this prior distribution, the AMP procedure utilises $\mathbf{Y}$ to find {the estimate} $\hat{\mathbf{X}}$ that minimizes the mean squared error (MSE)
\begin{equation}
    \text{MSE}= \mathbb{E}_{\mathbf{X},\mathbf{Y}}\norm{\hat{\mathbf{X}}(\mathbf{Y})- \mathbf{X}}_2^2.
    \label{mseAMP}
\end{equation}

Algorithm~\ref{algoAMP} presents details of {the} minimization steps of \eqref{mseAMP}. Line~1 is the initialization step, where the estimate $\hat{\mathbf{X}}^t \in \mathbb{C}^{NQ \times M}$ and residual $\mathbf{R}^t \in \mathbb{C}^{T \times M}$ at the iteration index $t$ are set to zero. The iterative procedure is repeated between Lines 2 and 6 until convergence. Line 3 computes the current estimate $\hat{\mathbf{x}}_{nq}^{t+1}$ using the MMSE denoising function defined as \cite{senel2018grant,wei2018joint}
\begin{equation}
        \eta(\hat{\mathbf{x}}_{nq}^t,\mathbf{\Sigma}) = \frac{\gamma_n(\gamma_n\mathbf{I}+\mathbf{\Sigma})^{-1}\hat{\mathbf{x}}_{nq}^t}{1+\frac{1-\xi}{\xi}	\eth(\hat{\mathbf{x}}_{nq}^t) \mathrm{det}\left(\mathbf{I}+\gamma_n\mathbf{\Sigma}^{-1} \right)   },
       \label{eta}
\end{equation}
where $\eth (\hat{\mathbf{x}}_{nq}^t)= \mathrm{exp}(-(\hat{\mathbf{x}}_{nq}^t)\herm\mathbf{\Sigma}^{-1}(\hat{\mathbf{x}}_{nq}^t)+ (\hat{\mathbf{x}}_{nq}^t)\herm(\gamma_n \mathbf{I}+\mathbf{\Sigma})^{-1}\hat{\mathbf{x}}_{nq}^t)$. Line 4 computes the residual and  $\eta'$ is the element-wise first order derivative of $\eta$, hence  it is a diagonal matrix for each $\mathbf{x}_{nq}$. The $\langle \rangle$ operator computes the average over all the $NQ$ matrices. 
It can be observed that~(\ref{eta}) depends on $ \mathbf{\Sigma}$, which is referred to as the state evolution \cite{wei2018joint}. This is a metric that measures the average MSE in Line 5, where $\mathbf{e} =\eta(\mathbf{x}_{nq}+(\mathbf{\Sigma}^{t})^{\frac{1}{2}}\mathbf{\bm{\nu}})-\mathbf{x}_{nq}$, and ${\bm{\nu}\sim\mathcal{CN}(\mathbf{0},\mathbf{I}) \in \mathbb{C}^{M \times 1}}$ is a random error independent of $\mathbf{x}_{nq}$. 

\begin{algorithm}[!t]
  \KwIn{$\mathbf{Y}$, $\Delta$, $\eta$}
 % \KwOut{$\hat{\mathbf{X}}^t$}
  Initialization: $\hat{\mathbf{X}}^0 = \mathbf{0}$, $\mathbf{R}^0=\mathbf{Y}$, $t= 0$\\\
    \Repeat{$   \norm{\mathbf{R}^{t+1}-\mathbf{R}^t}_{F} < \Delta$}{%
      %\label{stepB}
      
   $\hat{\mathbf{x}}_{nq}^{t+1} = \eta((\mathbf{R}^t)\herm\mathbf{s}_n^q +\hat{\mathbf{x}}_{nq}^t)$, $\forall n, \forall q$
   
    $\mathbf{R}^{t+1}={\mathbf{Y}}-\mathbf{S}\mathbf{X}^{t+1}+\frac{NQ}{T}\langle \eta'((\mathbf{R}^t)\herm\mathbf{s}_n^q+\hat{\mathbf{x}}_{nq}^t ) \rangle$\
       
        $\mathbf{\Sigma}^{t+1}=\frac{\sigma^2\mathbf{I}}{T\rho_{n}^{\mathrm{UL}}} +\frac{NQ}{T}\mathbb{E}(\mathbf{e}\mathbf{e}\herm)$\

      % $\hat{\mathbf{X}} =[\mathbf{x}_1^{t+1}, \mathbf{x}_2^{t+1},\cdots,\mathbf{x}_N^{t+1}]$\
      
       $t = t + 1$\
       
    }
     \KwOut{$\hat{\mathbf{X}} = \mathbf{X}^t$}
  \caption{AMP decoding}
  \label{algoAMP}
\end{algorithm}

%%%%%%%%%%%%%%%%%%%%%%%%%%%%%%%%%
\subsection{Solution via ADMM based $\ell_{2,1}$-minimization}
\label{solutionADMM}
The design of the AMP denoising function  described in Section \ref{solutionAMP} depends heavily on the fact that the large scale fading coefficients $\gamma_n,~ n=1,\cdots,N$, and activation probability $\epsilon$ are known at the BS. However, if such information is  not available at the BS, for instance, when a device is not active for a long time, mixed-norm minimization approaches \cite{steffens2018compact} {constitute} a viable solution as they are insensitive to the prior distribution. In practice, the most commonly used mixed-norm minimization approach is the $\ell_{2,1}$-norm minimization, given as  
\begin{equation}
 \operatorname*{minimize}_{\mathbf{X}} \frac{1}{2}\norm{\mathbf{Y}-\mathbf{S}\mathbf{X}}_{F}^2 + \mu \norm{\mathbf{X}}_{2,1},
   \label{l_2,1}
\end{equation}
where ${\mathbf{X}_{2,1} = \norm{\norm{\mathbf{x}_{11}}_2,\cdots,\norm{\mathbf{x}_{NQ}}_2}_1}$. Note that the problem  \eqref{l_2,1} is convex and can
be solved optimally using standard convex optimization
method {and solvers} such as cvx \cite{boyd2004convex}. However, since the expected number of devices is high in MTC, this approach is rendered highly inefficient.

Motivated by the need for a computationally efficient solution to \eqref{l_2,1}, we next present an iterative solution via an ADMM framework, where the update equations are in closed-form. First, we introduce an auxiliary variable ${\mathbf{Z}\in \mathbb{C}^{NQ\times M}}$ and rewrite  (\ref{l_2,1}) as
\begin{subequations}\label{P1}
    \begin{alignat}{2}
    &\underset{\mathbf{X},\mathbf{Z}}{\mathrm{minimize}} \quad && \frac{1}{2}\norm{\mathbf{Y}-\mathbf{S}\mathbf{Z}}_{F}^2+ \mu \norm{\mathbf{X}}_{2,1} \label{eqcostw2} \\
    &\text{subject to} 
     \quad && \mathbf{Z} = \mathbf{X}.\label{constrain12} 
    \end{alignat}
    \label{admmfORM}
\end{subequations}
Subsequently, the augmented Lagrangian associated with problem \eqref{admmfORM} is given as
\begin{equation}
    \mathcal{L}(\mathbf{X},\mathbf{Z},\mathbf{\Lambda}) \!=\! \frac{1}{2}\norm{\mathbf{Y}\!-\!\mathbf{SZ}}_{F}^{2} \!+\! \mu \norm{\mathbf{X}}_{2,1}\!+\!\frac{\rho}{2}\norm{\mathbf{S}\!-\!\mathbf{Z}\!+\!\frac{\mathbf{\Lambda}}{\rho}}_{F}^{2},
\end{equation}
where $\mathbf{\Lambda}=[\boldsymbol{\lambda}_{1,1},\cdots,\boldsymbol{\lambda}_{1,Q},\cdots,\boldsymbol{\lambda}_{N,1},\cdots, \boldsymbol{\lambda}_{N,Q}]\tran\in \mathbb{C}^{NQ \times M}$ is the matrix of ADMM dual variables and  $\rho$ is a positive parameter that controls the convergence speed of ADMM \cite{boyd2011distributed}. The solution to   problem \eqref{admmfORM}  via ADMM is achieved by  sequentially updating   $(\mathbf{Z},\mathbf{X},\mathbf{\Lambda})$ as follows\begin{equation}
 \mathbf{Z}^{t+1}:=\min_{\mathbf{Z}}\frac{1}{2}\|
 \mathbf{Y}-\mathbf{S}\mathbf{Z}\|_{\mathrm{F}}^2+ \frac{\rho}{2} \Vert   \mathbf{X}^{t} - \mathbf{Z} +\frac{1}{\rho}\mathbf{\Lambda}^{t} \|_{\mathrm{F}}^2, 
\label{eq::z(t+1)}
\end{equation}
\begin{equation}
 \mathbf{X}^{t+1}:=\min_{\mathbf{X}}   \mu \norm{\mathbf{X}}_{2,1}+\frac{\rho}{2} \|\mathbf{X}-\mathbf{Z}^{t+1}+ \frac{1}{\rho}\mathbf{\Lambda}^{t} \|_{\mathrm{F}}^2,
\label{eq::x(t+1)}
\end{equation}
  \begin{equation}
\mathbf{\Lambda}^{t+1} := \mathbf{\Lambda}^{t}+\rho\big(\mathbf{X}^{t+1}  -\mathbf{Z}^{t+1} \big).
\label{eq::lambda(t+1)}
\end{equation}The ADMM solution is detailed in \textbf{Algorithm~\ref{algoADMM}}. Step 3 is the update for $\mathbf{Z}$. It is is obtained by setting the derivative of the convex objective function \eqref{eq::z(t+1)} to zero and solving for $\mathbf{Z}$. The $\mathbf{X}$-update in Line 5 is computed using the soft-thresholding operator \cite{boyd2011distributed}, by decoupling \eqref{eq::x(t+1)} into $NQ$ convex sub-problems (for $n=1,\cdots,N$ and $q=1,\cdots,Q$) as
\begin{equation}
\begin{array}{cc}
      \mathbf{x}_{nq}^{t+1}\!:=\!\min_{\mathbf{x}_{nq}}   \mu  \Vert  \mathbf{x}_{nq}\|_2\!+\!\frac{\rho}{2} \|\mathbf{x}_{nq}\!-\! \mathbf{z}_{nq}^{t+1}+ \frac{1}{\rho}\boldsymbol{\lambda}_{nq}^{t} \|_{\mathrm{F}}^2.
\end{array} 
\label{eq::xw(t+1)}
\end{equation}

\begin{algorithm}[t]
  \KwIn{$\mathbf{Y}$, $\Delta$}
 
  Initialisation: $\mathbf{X}^0 = \mathbf{0}$, $\mathbf{Z}^0=\mathbf{0}$, $\mathbf{\Lambda}^0=\mathbf{0}$, $t= 0$\\\
    \Repeat{$\norm{\mathbf{Z}^{t+1}-\mathbf{Z}^t}_{F} < \Delta$}{%
      \label{stepA}
      
   $\mathbf{Z}^{t+1} =\left( \rho\mathbf{X}^{t} + \mathbf{\Lambda}^{t} + \mathbf{Y}\tran\mathbf{S} \right)\left( \mathbf{S}\tran\mathbf{S} + \rho\mathbf{I}_{NQ}\right)^{-1}$\
   
   $\mathbf{c}_{nq}^{t} = \mathbf{z}_{nq}^{t+1} - \frac{1}{\rho}\boldsymbol{\lambda}_{nq}$,$\forall q, \forall n$\
   
    $\mathbf{x}_{nq}^{t+1} = \frac{\max\left\{\mathbf{0},\norm{\mathbf{c}_{nq}^{t}}_{2}-\frac{\mu}{\rho}\right\}\mathbf{c}_{nq}^{t}}{\norm{\mathbf{c}_{nq}^t}_{2}}$,$\forall q, \forall n$\

    $\mathbf{\Lambda}^{t+1} = \mathbf{\Lambda}^{t} + \rho \left(\mathbf{X}^{t+1} - \mathbf{Z}^{t+1} \right)$\

      % $\hat{\mathbf{X}} =[\mathbf{x}_1^{t+1}, \mathbf{x}_2^{t+1},\cdots,\mathbf{x}_N^{t+1}]$\
      
       $t = t + 1$\
    }
    \KwOut{$\hat{\mathbf{X}}= \mathbf{Z}^t$ }
  \caption{ADMM decoding}
  \label{algoADMM}
\end{algorithm}

\subsection{Solution via Expectation-Maximization Sparse Bayesian Learning}
\label{SBLsection}

{
Sparse Bayesian learning (SBL), an important family of Bayesian SSR algorithms, recovers a sparse vector through a hierarchical signal prior model. The key is to introduce real-valued hyperparameters, $\bm{\tilde{\alpha}} = \{\alpha_{nq}\gamma_n\}_{n\in \mathcal{N},q =1,\cdots,Q}$, which represent the (co)variance of each  Gaussian channel $\{\tilde{\mathbf{g}}_n\}$. The main task in SBL is to estimate the sparse vector $\bm{\tilde{\alpha}}$, which consequently reveals the underlying sequence transmission indicator $\bm{\alpha}$. In SBL, for a given $\bm{\tilde{\alpha}}$, the signal estimate is given as the MMSE estimate \cite{zhang2011sparse}, ${\hat{\mathbf{X}}=\mathbb{E}[\mathbf{X}|\mathbf{Y},\bm{\tilde{\alpha}}]}$, which is linear due to Gaussianity (see Line 3 in {Algorithm}~\ref{algoSBL}); here, $\tilde{\mathbf{\Gamma}} = \mathrm{diag}\{\bm{\tilde{\alpha}}\}$. The estimate for $\bm{\tilde{\alpha}}$ is found via the Type II maximum a posteriori probability (MAP) estimation as
%\begin{equation}\label{eq_SBL_cost}
%\begin{array}{ll}
% \bm{\tilde{\alpha}} &\hspace{-2mm}=   \underset{\bm{\tilde{\alpha}}}{\mathrm{argmax}}~\mathrm{log}\,p( \bm{\tilde{\alpha}}|\mathbf{Y}) \\  
 %&\hspace{-2mm}=   \underset{\bm{\tilde{\alpha}}}{\mathrm{argmin}}~T\mathrm{log}\,|\bm{\Sigma}_{\mathbf{y}}| + \Tr\left(\bm{\Sigma}_{\mathbf{y}}^{-1}\mathbf{Y}\mathbf{Y}\herm     \right) - p( \bm{\tilde{\alpha}}),
%\end{array}
%\end{equation}

\begin{align}
\bm{\tilde{\alpha}} &\hspace{-2mm}=   \underset{\bm{\tilde{\alpha}}}{\mathrm{argmax}}~\mathrm{log}\,p( \bm{\tilde{\alpha}}|\mathbf{Y})\nonumber \\  
 &\hspace{-2mm}=   \underset{\bm{\tilde{\alpha}}}{\mathrm{argmin}}~T\mathrm{log}\,|\bm{\Sigma}_{\mathbf{y}}| + \Tr\left(\bm{\Sigma}_{\mathbf{y}}^{-1}\mathbf{Y}\mathbf{Y}\herm     \right) - p( \bm{\tilde{\alpha}}),
   \label{eq_SBL_cost} 
\end{align}where $\bm{\Sigma}_{\mathbf{y}} = \sigma^2\mathbf{I} + \mathbf{S}\tilde{\mathbf{\Gamma}}\mathbf{S}\herm $ is the covariance matrix\footnote{This covariance matrix should be dependent on the accuracy of SIC through $\bm{w}_d$. Since it is impractical to know the accuracy of SIC, we assume the best case scenario of SIC.} of the signal remaining after SIC of eMBB signals; $p(\bm{\tilde{\alpha}})$ is the sparsity-promoting prior distribution of the hyperparmeter. In this work, non-informative $p( \bm{\tilde{\alpha}})$ is used, hence it has no impact in the minimization.}

% % a To provide this solution, we first establish the posterior distribution, which is defined by  %{To acquire the solution via EM-SBL, we define the posterior of the parameters of interest, i.e., the sparsity promoting hyper-parameter and the channels}
% \begin{align}
%     \mathcal{P}(\tilde{\mathbf{X}},\tilde{\mathbf{\Gamma}}|\mathbf{Y}) \propto   \mathcal{P}(\mathbf{Y}|\tilde{\mathbf{\Gamma}},\mathbf{X})   \mathcal{P}(\mathbf{X}|\tilde{\mathbf{\Gamma}})\mathcal{P}(\tilde{\mathbf{\Gamma}}), 
%     \label{EMSBL}
% \end{align}
% where $\tilde{\mathbf{X}} = \tilde{\mathbf{G}}$ and $\tilde{\mathbf{\Gamma}}= \mathbf{D}\mathrm{diag}(\gamma_1,\cdots,\gamma_{NQ})$ is the sparsity promoting parameter. From \eqref{EMSBL}, the marginal distribution and the estimate of $\tilde{\mathbf{\Gamma}}$ can be computed by
% \begin{align}
%   \mathcal{P}(\tilde{\mathbf{\Gamma}}|\mathbf{Y}) = \int\mathcal{P}(\mathbf{Y}|\tilde{\mathbf{\Gamma}},\mathbf{X})   \mathcal{P}(\mathbf{X}|\tilde{\mathbf{\Gamma}})\mathcal{P}(\tilde{\mathbf{\Gamma}})d\mathbf{X},\\ 
%   \tilde{\mathbf{\Gamma}} = \text{argmax}_{\tilde{\mathbf{\Gamma}}}  \mathcal{P}(\tilde{\mathbf{\Gamma}}|\mathbf{Y})
%   \label{sblcosts}
% \end{align}

We solve \eqref{eq_SBL_cost} via expectation-maximization (EM) \cite{zhang2011sparse}, hence the name EM-SBL. The iterative solution is presented in \textbf{Algorithm}~\ref{algoSBL}.
%In Line 1, the algorithm is initialized. 
Line 3 updates the estimate of $\mathbf{X}$,  whereas Line 5 updates the sparsity-promoting hyperparameter.

\begin{algorithm}[!t]
  \KwIn{$\mathbf{Y}$, $\Delta$, ${\sigma^2}$}
 % \KwOut{$\hat{\mathbf{X}}^t$}
  Initialization: $\mathbf{X}^0 = \mathbf{0}$, $\bm{\tilde{\alpha}} = \bm{1}_{NQ \times 1}$, $t= 0$\\\
    \Repeat{$   \frac{\norm{\mathbf{X}^{t+1}-\mathbf{X}^t}_{F}}{\norm{\mathbf{X}^t}_{F}} < \Delta$}{%
      %\label{stepB}
      $\mathbf{X}^t = \tilde{\mathbf{\Gamma}}\mathbf{S}\tran\left(\sigma^2\mathbf{I} + \mathbf{S}\tilde{\mathbf{\Gamma}}\mathbf{S}\tran\right)^{-1}\mathbf{Y}  $\
     
    $\mathbf{F}^t = \left(\tilde{\mathbf{\Gamma}}^{-1} + \frac{1}{\sigma^2}\mathbf{S}\tran\mathbf{S}\right)^{-1}  $\
    
        $\bm{\tilde{\alpha}} = \frac{1}{M}\norm{\mathbf{x}_{nq}^t}_{2}^{2} + \mathbf{F}_{ii}, i = \{1,2,\cdots,NQ\},\forall n, \forall q $\
     
       $t = t + 1$\
       
    }
     \KwOut{$\hat{\mathbf{X}} = \mathbf{X}^t$}
  \caption{EM-SBL decoding}
  \label{algoSBL}
\end{algorithm}

\subsection{Solution Via SOMP}
  \label{OMPsection}
  The SOMP algorithm \cite{cai2011orthogonal,pati1993orthogonal} identifies the non-zeros entries of the sparse matrix $\mathbf{X}$ by iteratively selecting  the columns of $\mathbf{S}$ that have the highest correlation with the residual $\mathbf{R}$ at each step. It is presented as \textbf{Algorithm}~\ref{algoOMP}. Initially, the residual $\mathbf{R}$ is set to be equal to the version of $\mathbf{Y}$  that remains after SIC of the eMBB signals. Line 3 computes the correlation between $\mathbf{S}$ and the current residual, while Line 4 selects the column of $\mathbf{S}$  that has the highest correlation with $\mathbf{R}$. These columns are accumulated in $\mathcal{H}$ in Line 5. To avoid repetition, in Line 7, the selected columns' contributions are removed from $\mathbf{Y}$, thus leaving a residual that is orthogonal to the previously selected columns.
\begin{algorithm}[!h]
  \KwIn{$\mathbf{Y}$, $\Delta$}
 % \KwOut{$\hat{\mathbf{X}}^t$}
  Initialization: $\hat{\mathbf{X}}^0 = \mathbf{0}$, $\mathbf{R}^0=\mathbf{Y}$, $t= 0$,$\mathcal{H}^{[0]} = \emptyset $\\\
    \Repeat{$   \frac{\norm{\mathbf{X}^{t}-\mathbf{X}^{t-1}}_{F}}{\norm{\mathbf{X}^{t}}_{F}} < \Delta$}{%
      %\label{stepB}

       $\mathbf{\mathcal{G}}^{t} = \mathbf{S}\herm \mathbf{R}^{t-1}$\\
     
        $j^{t} = \displaystyle\text{argmax}_{j}   \left\{\frac{\norm{\mathbf{\mathcal{G}}_{j}^{t}}_{1}}{\norm{\mathbf{S}_j}_{2}} \right\} $\
         % $T^{[t]} = T^{i-1}\cup j^{t} \mathbf{R}^{[t-1]}$\
          
          $\mathcal{H}^{t} = \mathcal{H}^{t-1}\cup j^{t} $\\
      % $j^{[t]} = \text{argmax}_{j}\frac{|g^{[t]}|}{\norm{\mathbf{A}_j}_{2}}$\\
           $\mathbf{X}_{[\mathcal{H}^{t}]}^{t} =\mathbf{S}_{[\mathcal{H}^{t}]}^{\dagger} \mathbf{Y} $\\
        $\mathbf{R}^{t} = \mathbf{Y} - \mathbf{S}\mathbf{X}^{t}$\
        
       $t = t + 1$\
       
    }
     \KwOut{$\hat{\mathbf{X}} = \mathbf{X}^t$}
  \caption{SOMP decoding}
  \label{algoOMP}
\end{algorithm}
 %\textcolor{blue}{ \subsection{Complexity Analysis}
% A typical 5GB cellular network is composed of a massive number of MTDs and a large number of eMBB devices. Evidently, computational complexity of the algorithms running at the BS is of critical importance as it influences power consumption, which essentially contributes to the carbon footprint. In table \ref{complexAnalysis} we provide a  summary of these complexities. Algorithm~1, which is used for optimizing sequences of the MTDs is of high complexity. However, sequences can be generated once and stored at the BS, thus reducing power consumption. From the SSR algorithms, it can be noted that both AMP and OMP are of low complexity, followed by the $\ell_{2,1}$-ADMM. On the other hand, it can be observed that the EM-SBL is of high complexity.}
 \begin{table*}[t!]
\centering
\caption{Computational complexity for decoding using different SSR techniques}

{\begin{tabular}{c||c||c} 
 \hline
  Algorithm & Number of complex operations in each iteration & $\mathcal{O}(\cdot)$ \\ [0.5ex] 
 \hline\hline
AMP & $NQTM + NQTM + NMQ + M $ & $\mathcal{O}(NQTM)$\\ 
$\ell_{2,1}$-ADMM & $MTNQ + MN^2Q^2$ & $\mathcal{O}(MN^2Q^2 + MTNQ)$\\
EM-SBL  & $2M^3N^2Q^2 T + M^2T^2 + NQM^2T$ & $\mathcal{O}(N^2Q^2 M^3T + M^2T^2)$ \\
 SOMP & $(2T + 1)MNQ + TM^2 + (M +1)T$ & $\mathcal{O}(TMNQ)$ \\ [1ex] 
 \hline
\end{tabular}}
\label{complexAnalysis}
\end{table*}

%%%%%%%%%%%%%%%%%%%%%%%%%%%%%%%%%%%%%%%%
\section{Numerical Results and discussions}
In this section, we present numerical results to show {the} performance of our proposed framework in terms of the average outage probability, sequence detection accuracy, and channel estimation accuracy.

\subsection{Simulation Setup}
%In the experimental setup, 
We consider a single-cell network of radius $250$ m, where the BS serves $N = 1000$ MTDs and 
$E = 4$ eMBB devices, all randomly placed in the cell area. The path-loss for each device is computed as a function of the device's distance $d_i$, $i\in \{\mathcal{E}\cup \mathcal{N}\}$, from the BS. We adopt a power control mechanism which ensures that the average received power is the same for all the transmitted messages, irrespective of the device' s position. Since the large scale parameters are known at the BS, each MTD transmits with power
\begin{equation}
    p_{n}^{\text{UL}} = \frac{p^{\mathrm{max}}\gamma_{\mathrm{min}}}{\gamma_n},
\end{equation}
\noindent where $\gamma_{\mathrm{min}}$ is the large scale effect at the edge of the cell and $p^{\mathrm{max}}$ is the maximum allowable power for each MTD. Then, the average received SNR corresponding to the $n$-th MTD is given by
\begin{equation}
    \text{SNR}_{n}=\frac{p^{\mathrm{max}}\gamma_{\mathrm{min}}}{\sigma^2}.
    \label{snrValues}
\end{equation}
With a similar analogy, the average received SNR for the $e$-th eMBB device is given by
\begin{equation}
    \text{SNR}_{e}=\frac{\rho^{\mathrm{max}}\beta_{\mathrm{min}}}{\sigma^2}.
    \label{snrEmbb}
\end{equation}The numerical results are obtained over $10^3$ channel realizations.
 A summary of simulation parameters  is provided in Table~\ref{simulationPar}.

 \begin{table}
    \centering
  \caption{Simulation parameters}
 \begin{tabular}{  m{16em}  m{10em} } 
\hline
\textbf{Parameter} & \textbf{Value}\\ 
\hline

Path loss for distance (km) &  $130 + 37.6 \mathrm{log}_{10}(d_i)$ dB   \\

Bandwidth  & $20$ MHz  \\ 

Noise power ($\sigma^2$) & $2 \times 10^{-13}$ W \\ 

Maximum uplink transmission power& $0.1$ W \\ 

Number of bits $(g)$& 128 \\ 
Symbol duration $(T_s)$ &$16~\mu$s \\ 
eMBB pilot length $(L)$& 32 \\ 
Coherence interval $(T)$ & $256$ \\ 
Cell radius & $250$ m \\ 
Number of MTDs $(N)$ & $1000$ \\
Number of eMBB devices $ (E)$ & $4$ \\
Number of BS antennas $(M)$ & $32$ \\ 
Activation probability $(\epsilon)$ & $0.01$ \\ 
$\text{SNR}_{e}$ & $25$ dB \\ 
$\text{SNR}_{n}$ & $5$ dB\\ 
Error tolerance $(\Delta)$ & $10^{-4}$ \\ 
Number of messages $(Q)$ & $2$ \\ 
Collision probability $(\chi)$ & $10^{-6}$ \\
\hline
\end{tabular}
\label{simulationPar}
\end{table}
 \label{results}
%\subsection{Performance Analysis}

\subsection{Performance Metrics}
To evaluate the eMBB performance achieved under the proposed solutions, we analyze the {average} outage probability of the eMBB devices. We mimic a practical procedure by assuming that an $e$-th eMBB device is successfully decoded if $\mathrm{\Gamma}_e$ is greater than a certain threshold \cite{xia2018outage}. We define this threshold as $2^r-1$, where\begin{equation}
     r = \frac{b}{(T-L)T_{s}}
     \label{rateVal}
 \end{equation}is the transmit data rate in bits per channel use (bpcu), $b$ and $T_{s}$ are the number of bits and symbol time, respectively. Therefore, the average outage probability\footnote{Given that the eMBB devices are exposed to similar conditions, their average outage probability is the same as for one eMBB device.} of the eMBB devices is defined by
 \begin{equation}
    P_{out} = \mathrm{Pr}(\mathrm{\Gamma}_e<2^r-1).\label{outageAnalysis}
\end{equation}
For the MTC network, we use {the} receiver operating characteristics (ROC) to quantify the inherent trade-offs related to the probability of miss detection (PMD) and probability of false alarm (PFA). It is worth highlighting that PMD and PFA are computed at {the} sequence level. For example PMD is the probability of missing a message rather than an active MTD as conventionally done in computing ROC \cite{senel2018grant}. Numerically, the PMD and PFA are respectively computed as\begin{subequations}\begin{align}
\text{PMD} =  \mathbb{E}\left( \frac{\sum_{i=1}^{NQ}\text{max}(0,\alpha_i - \hat{\alpha_i})}{|\mathcal{K}|} \right) \\ \text{PFA} = \mathbb{E}\left( \frac{\sum_{i=1}^{NQ}\text{max}(0,\hat{\alpha}_i-\alpha_i  )}{{NQ}-|\mathcal{K}|} \right).
%\nonumber
\end{align}
\end{subequations}
We also evaluate the normalized mean squared error (NMSE)  of the channel estimates. This measures the relative error in the channel estimate and hence defined for the eMBB devices and MTDs as
\begin{equation}
  \text{NMSE}_{i}=  \mathbb{E}_{i} \left(\frac{\norm{\mathbf{h}_{i} - \hat{\mathbf{h}}_{i}}^2}{\norm{\mathbf{h}_{i} }^2}\right),  i\in \mathcal{E}\cup\mathcal{K}.
  \label{channelMtds}
\end{equation}
Even though non-coherent detection does not require explicit CSI, we found it necessary to reveal the explicit performance of the SSR algorithms in the identification of the non-zero rows of $\mathbf{X}$ by using \eqref{channelMtds}. This serves as a litmus test because performance achieved using this metric is not subject to external settings. On the other hand, detection and decoding evaluated by PMD and PFA are normally subject to how well a threshold is set.

%\textcolor{blue}{results at SNR of $30$ dB show an NMSE of $10^{-8}$ for $L=128$ and $10^{-7}$ for $L = 64$ when using the proposed pilot generation strategy. On the other hand, results for the SNR of $30$ dB show an NMSE of $10^{-2}$ for $L = 128$ and $10^{-1}$ for $L = 64$ when using Gaussian pilots. NMSE of $10^0$ is reported for the different pilot generation strategies at $-20$ dB.}

%%%%%%%%%%%%%
\subsection{Performance of the eMBB}
\label{sparsityeffect}
\begin{figure}
\begin{subfigure}{0.5\textwidth}
  \centering
  \includegraphics[width=0.9\linewidth]{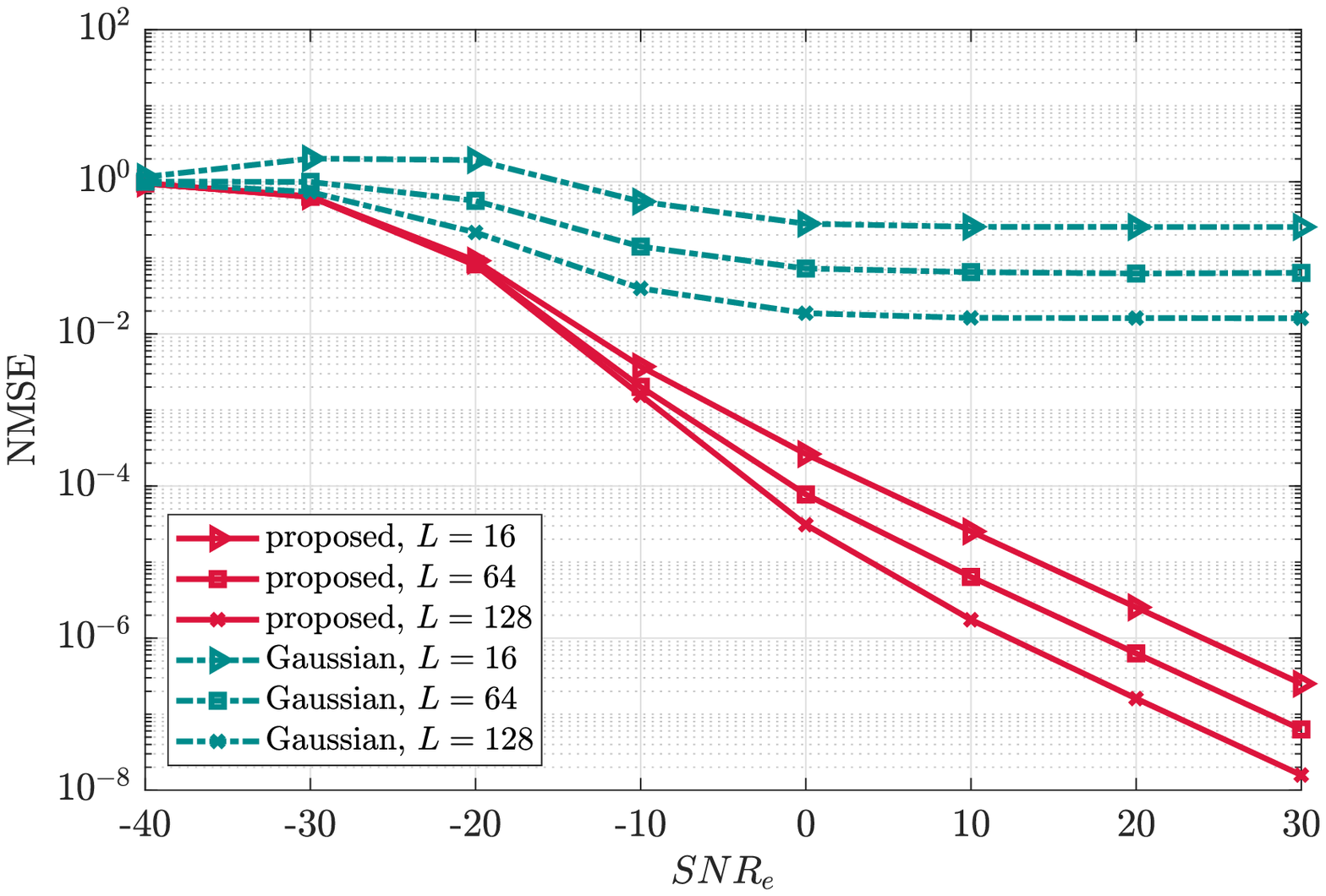}  
 % \caption{ \textcolor{blue}{Average channel estimation NMSE for different pilot lengths vs. SNR of the eMBB device.}}
  \label{embbCsiEstimate}
\end{subfigure}
%\hfill
\
\begin{subfigure}{0.5\textwidth}
  \centering
  \includegraphics[width=0.9\linewidth]{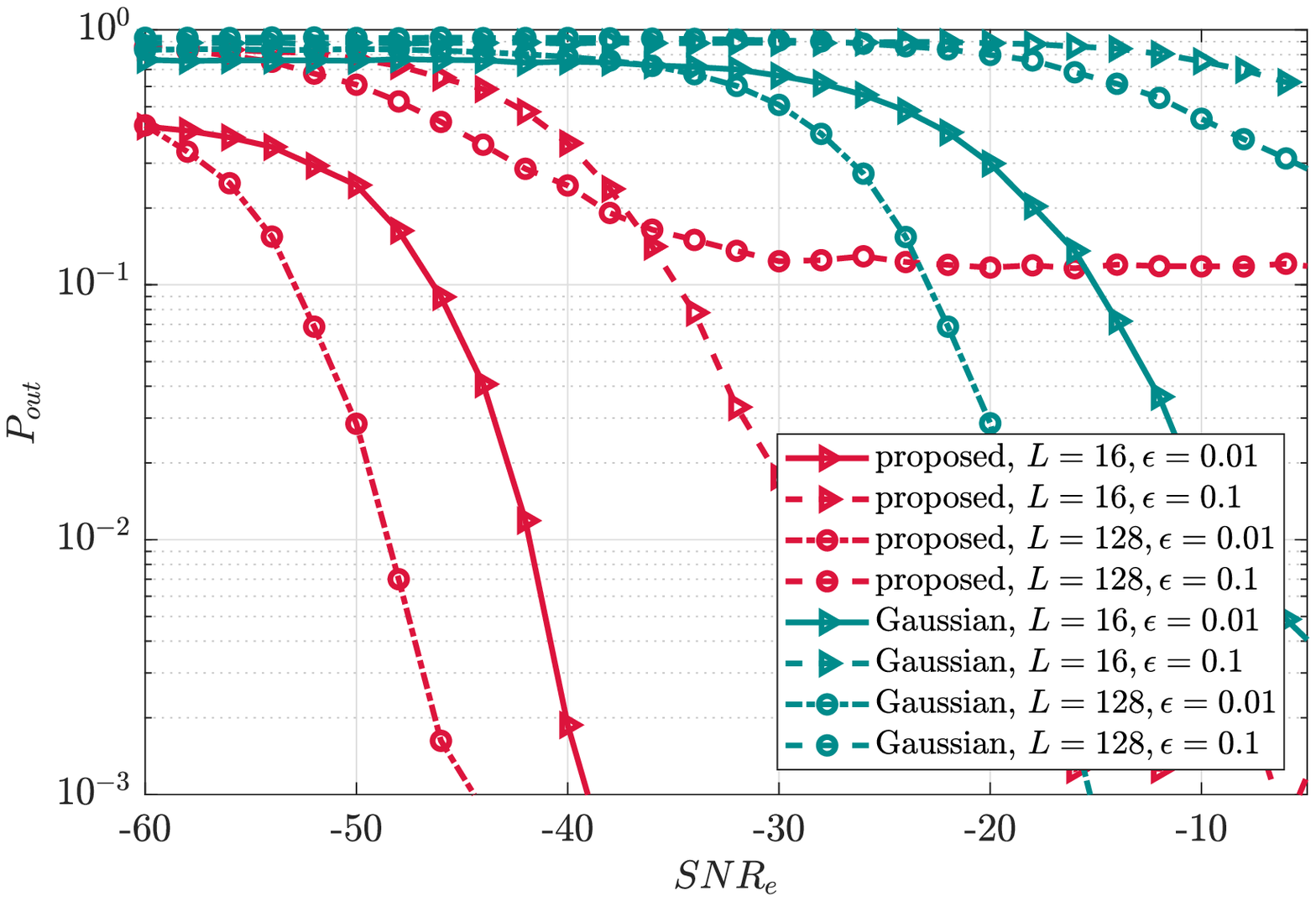}  
  %\caption{\textcolor{blue}{average eMBB outage probability for different activation probabilities and different pilot lengths}}
  \label{outageEmbb}
\end{subfigure}
%\vspace{-3mm}
\caption{(a) Average channel estimation NMSE as a function of $\mathrm{SNR}_{e}$ and $L$ (left) (b) Average eMBB outage probability as a function of $\mathrm{SNR}_{e}$ and $\epsilon$ for different $L$ (right).}
\label{ActivitySparsePilots}%\vspace{-7mm}
\end{figure}

Fig.~\ref{ActivitySparsePilots} shows the performance of the eMBB devices in the presence of MTC traffic. Specifically, Fig. \ref{ActivitySparsePilots}(a) shows the channel estimate performance as a function of the average SNR and the pilot length. We have used Gaussian pilot sequences \cite{senel2018grant,wei2018joint,ahn2019ep,chen2019covariance} as a benchmark to show the superiority of our proposed framework. On average, results evince that increasing both the average SNR (by increasing $\rho_e$) and pilot lengths of the eMBB devices improves performance. These results are expected since an increase in pilot length corresponds to longer training, which reduces estimation errors. Similarly,  higher transmission powers encourage predominance of the signal over low power interference and noise, thus, improving performance.
%\textcolor{red}{The above} results confirm that the proposed pilot generation strategy indeed makes it possible to acquire interference-free channel estimate of \textcolor{blue}{eMBB devices}.}
A superficial analysis of these results can give the impression that longer pilot length (for the eMBB devices) guarantee a favourable coexistence with the MTDs. However, it can be quite the opposite. To assess this, Fig.~\ref{ActivitySparsePilots}(b) shows the {average} outage probability of the eMBB devices as a function of the activation probabilities ($\epsilon$) of {the} MTDs and the pilot lengths of the eMBB devices. For fixed $L$, it is observable that higher outage probability occurs with $\epsilon = 0.1$ than with $\epsilon = 0.01$. Also, it can be noted that for fixed $\epsilon = 0.1$, the outage probability is higher for $L= 128$ than for $L = 16$. It is apparent that these two factors, i.e., increased activity and increased pilot lengths negatively impact the network performance by increasing the outage probability of the eMBB devices. Their impact can be interpreted using the necessary condition for decoding data from eMBB devices as detailed in \eqref{outageAnalysis}. Explicitly, we note that an increased $\epsilon$ probabilistically increases $|\mathcal{K}|$, which increases interference and thus reduces the SINR ($\mathrm{\Gamma}_e$). Similarly, increasing $L$ leads to {a} higher transmission rate ($r$) and thus increases the threshold required for successfully decoding eMBB signals. Nevertheless, both lead to {the} violation of the required condition to successfully decode the data sent by eMBB devices. In turn, this reduces the SIC capability, which results in the eMBB signals causing stronger interference to the MTDs during the SSR phase. It is worth noting that such strong interference reduces the {BS'} ability to decode data from {the} MTDs, as will be seen next. For the remainder of this work, we only present results using the proposed pilot sequences as it performs better than the benchmark.

\subsection{Performance of the MTDs}
\begin{figure}[t]
    \centering
    \includegraphics[width=0.45\textwidth]{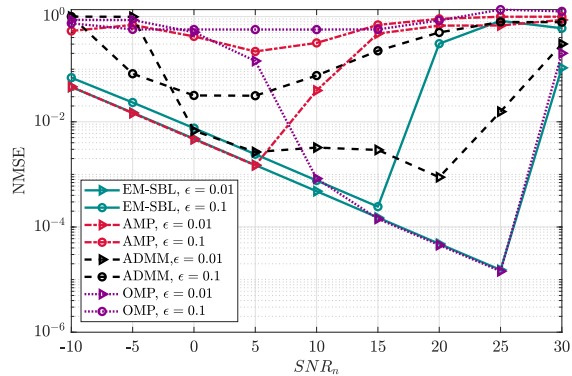}%\vspace{-2mm}
    \caption{Channel estimation NMSE for active MTDs as a function of $\mathrm{SNR}_{n}$ and $\epsilon$.}
    \label{OptimalSNR}%\vspace{-7mm}
\end{figure}

Fig.~\ref{OptimalSNR} shows the performance of decoding the MTDs in terms of the NMSE as a function of the average SNR and activation probability ($\epsilon$). In all the cases, we observe that the NMSE starts decreasing with increasing $\mathrm{SNR}_n$ until reaching a point  of inflection where the NMSE starts to increase. The inflection reveals the optimal point where messages from MTDs are successfully decoded using the SSR techniques. Consequently, this point is also an optimal point for favorable coexistence between MTDs and eMBB devices. It is also important to observe that the position of this point is influenced by both {average SNR} values and $\epsilon$. It can also be observed that EM-SBL and AMP generally have equal performance in low average SNR regimes and  also outperform OMP and $\ell_{2,1}$-ADMM due to the fact that they exploit more information from the measurements. In spite of that, it can be observed that AMP is more sensitive to the system configuration, which is consistent with the discussions pointed  out by the authors in \cite{jiang2022performance}.

In connection with {our discussions in} Section \ref{sparsityeffect}, high values of {average SNR} of {the} MTDs and $\epsilon$ increase the amount of interference to the eMBB signals, which leads to higher $P_{out}$ of the eMBB devices and failure to perform SIC. One of the interesting aspects of these results is that as $\epsilon$ increases, smaller values of average SNR of the MTC signals are preferred for a favourable coexistence. Unfortunately, reducing {average SNR} values of {the} MTDs leads to difficulty in decoding {their data}, which is generally undesirable. However, it is important to note that these results show that the proposed framework enables coexistence of the different services under certain configurations. To substantiate this, we subsequently present the ROC, hence showing inherent trade-offs between PMD and PFA based on the average SNR values and outage probabilities of the eMBB devices.
 
\begin{figure}[t]
    \centering
    \includegraphics[width=0.45\textwidth]{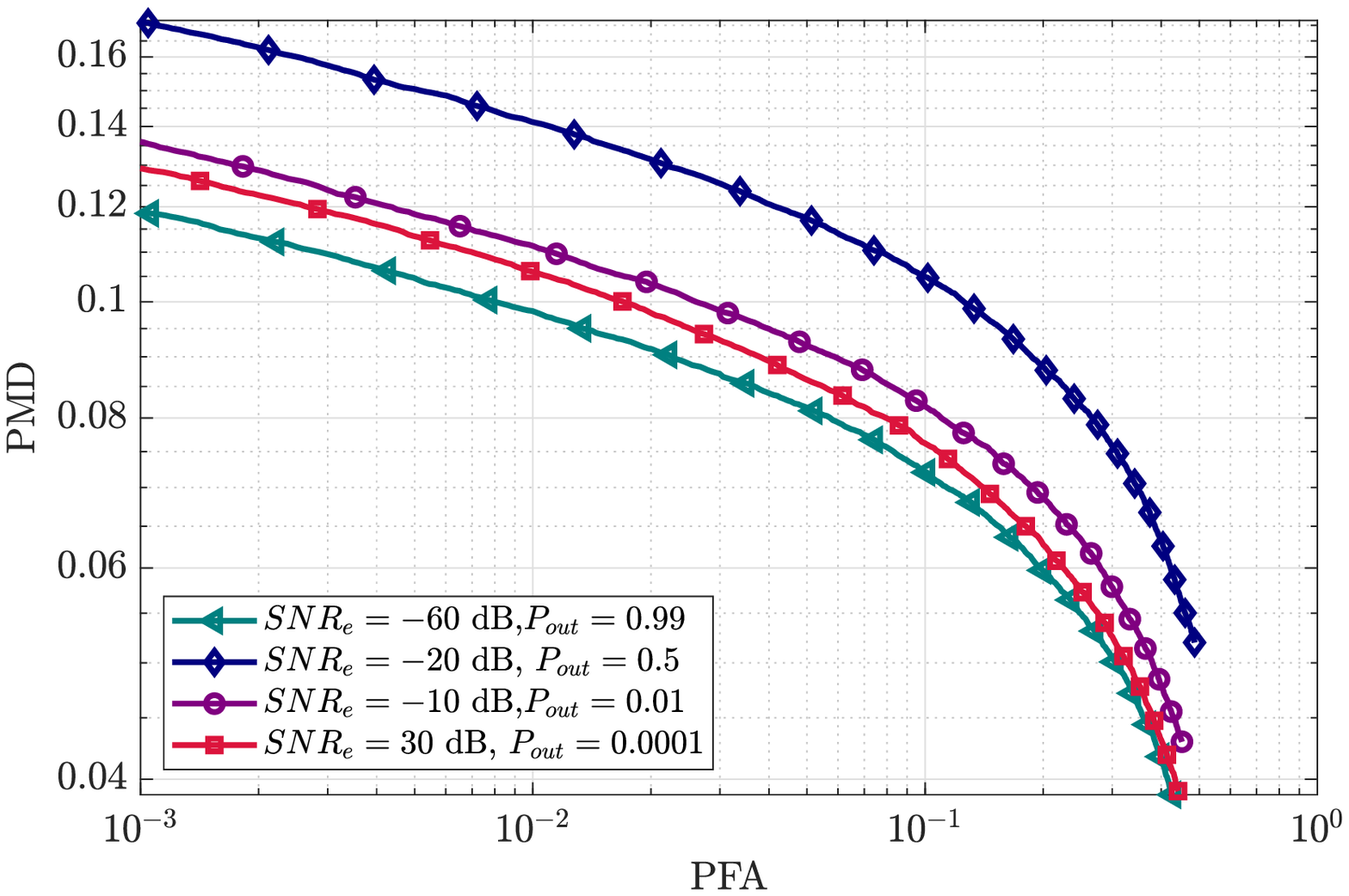}%\vspace{-2mm}
    \caption{ROC as a function of $\mathrm{SNR}_{e}$.}
    \label{rocFigsnr}%\vspace{-7mm}
\end{figure}

Fig.~\ref{rocFigsnr} shows ROC performance of AMP\footnote{The ROC of AMP in this region of operation is similar to that of EM-SBL, hence we only present these results.} with different {average SNR} values of the eMBB signals. The ROC curves in Fig.~\ref{rocFigsnr} can be used as a guide when choosing regions of operation for both MTDs and eMBB devices. To be explicit, we have included both the SNR levels of the eMBB devices hand in hand with the average outage probability for operating with these SNR values. It  can be observed that the data from MTDs is decoded more accurately when $\text{SNR}_{e} = -60$ dB than when it is $-20$ dB. This can be attributed to the fact that eMBB signals with average SNR of $-60$ dB do not cause significant interference to the MTDs and hence data from the MTDs can be decoded successfully. In spite of this (successful decoding), the average outage probability of the eMBB devices is high, as shown in Fig.~\ref{ActivitySparsePilots}(b), thus  rendering the coexistence unfavourable. However, operating eMBB devices at $\text{SNR}_{e} = -20$ dB degrades detection capabilities because this average SNR value does not guarantee that eMBB devices can be decoded yet causing significant interference to signals from the MTDs. At first instant, it can be inferred that as the average SNR of eMBB devices increases, their interference to the MTDs will be stronger and that the ability to decode data from the MTDs will degrade. Contrary to this line of analysis, it can be observed  that operating eMBB devices with average SNR of  $-10$ dB leads to an improved ability to decode the MTC signals as compared to when operating with $-20$ dB. This is so because average SNR of $-10$ dB guarantees that eMBB devices are decoded with high probability which makes it possible to perform SIC. Similarly, it can be observed that when eMBB devices are operated at $30$ dB, the performance approaches that achieved when operating eMBB devices at $-60$ dB. Intuitively, at average SNR of $30$ dB, there is little performance degradation in the detection of the MTDs data, although herein both eMBB and MTC are satisfactorily coexisting. These results confirm that the proposed joint coherent and non-coherent signal processing techniques come in handy in facilitating {the} coexistence of the diverse services in a 5GB network.

\begin{figure}[t]
    \centering
    \includegraphics[width=0.45\textwidth]{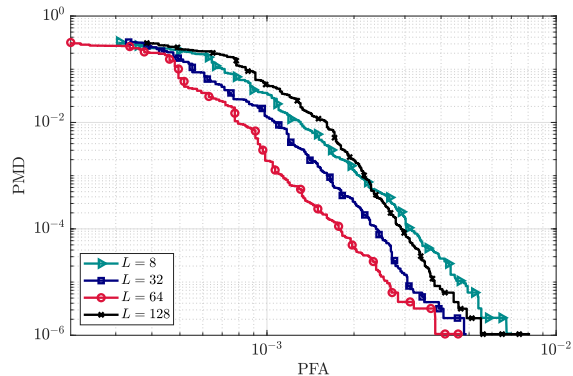}%\vspace{-2mm}
    \caption{ROC as a function of $L$, $\mathrm{SNR}_{e} = -10$ dB.}
    \label{pmdVsPFApilot}%\vspace{-7mm}
\end{figure}

%\textcolor{red}{. It is therefore important to note that any variation in $L$ impacts the \textcolor{blue}{eMBB devices} first and ultimately affecting MTDs through poor quality (channel estimate accuracy and SIC) processing of the \textcolor{blue}{data from eMBB devices}}
Fig.~\ref{pmdVsPFApilot} shows {the} ROC results for different pilot lengths of the eMBB devices. It is worth noting that due to non-coherent detection of {the MTC signals}, the size of $\mathbf{Y} \in \mathbb{C}^{T \times M}$ remains unchanged and hence the under-sampling ratio\footnote{Under-sampling ratio measures the number of measurements for a given vector length, e.g., $\frac{T}{NQ}$ means that there are $T$ available measurements to recover a vector of length $NQ$. Thus, a lower under-sampling ratio imposes a highly under-determined system.} also remains the same. From the results, it is clear that the BS' ability to decode the MTC signals degrades dismally with $L =8$ and $L= 128$, while the performance improves with $L=64$. Consistent with Fig.~\ref{ActivitySparsePilots}(b), $L = 128$ comes with high outage probability of the eMBB devices, which degrades the ability to decode the data from the MTDs. Intuitively, under the current configuration, our results show that there is favourable coexistence with an intermediate values of $L = 32$ and  $L = 64$. Therefore, it is important to note that while an increase in pilot length provides absolute gains in works such as \cite{senel2018grant,wei2018joint}, the situation is quite different in heterogeneous networks. These results confirm the need for our proposed framework. 

\begin{figure}[t]
    \centering
    \includegraphics[width=0.45\textwidth]{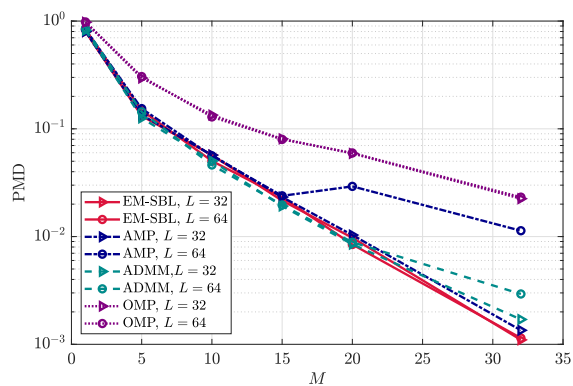}%\vspace{-2mm}
    \caption{PMD as a function of $M$ and $L$, $\mathrm{PFA} = 10^{-3}$}
    \label{AERantennasSizesf}%\vspace{-7mm}
\end{figure}

Fig.~\ref{AERantennasSizesf} shows the PMD (of all the SSR algorithms) as a function of $M$, for $\text{PFA} = 10^{-3}$. In general, it can be observed that for both $L= 32$ and $L = 64$, the PMD tends to decrease with the number of antennas at the BS. This improvement can be attributed to the fact that more antennas lead to increased diversity, which improves $\Gamma_e$ and also improves performance of the SSR algorithms \cite{senel2018grant}. From the figure it can also be noted that in general EM-SBL outperforms the other SSR algorithms. Next, we present the results {from} varying the size of the codebook for each MTD.

%\subsubsection{Impact of number of sequences for MTDs at the BS}

\begin{figure}[t]
    \centering
    \includegraphics[width=0.45\textwidth]{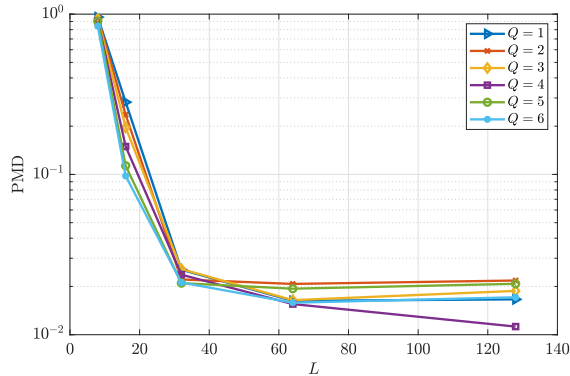}%\vspace{-2mm}
    \caption{PMD as a function of $L$ and $Q$, $\mathrm{PFA} = 10^{-3}$}
    \label{sequncesLenght}%\vspace{-7mm}
\end{figure}

Fig.~\ref{sequncesLenght} shows the performance in terms of the PMD as a function of $L$ and $Q$, for $\text{PFA} = 10^{-3}$ using EM-SBL. Here, we consider a worst case scenario where all the MTDs have the same set of messages. From the results, we observe that as the pilot length of the eMBB devices increases, the PMD decreases. This is mainly due to successful SIC and that the interference between the {data signal} of the MTDs and eMBB signals is lower. Particularly, it is worth noting that in the region where it is possible to decode the eMBB devices (e.g., average SNR of $30$ dB), as $L$ increases, $\mathbf{S}$ has minimal interference with ${\bm{\psi}_e}$. This improves the ability of the BS in detecting the data from the MTDs. To put this into perspective, we can consider the extreme case where $T=L$. In this case, the pilot sequences of the eMBB devices are completely orthogonal to the data from the MTDs and thus better detection capabilities are expected. On the other hand, with $L= 8$, each $\bm{\psi}_e$ is short, thus, each $\bm{\phi}_e$ imposes greater interference with $\mathbf{S}$, which ultimately degrades the BS's ability to decode the data from the MTDs. In spite of these, we observe a minimal impact of varying the codebook size. 

{ Fig.~\ref{htcVsMtd} shows the performance in terms of PMD as a function of $\epsilon$ and $E$ for ${\mathrm{PFA} = 10^{-3}}$. Results show that PMD increases with an increase in the number of the eMBB devices. Furthermore, performance degrades with increasing activity of the MTDs. This is in connection with the discussions in Section~\ref{sequence}. It is worth noting that as $E$ increases, there are $L-E$ orthogonal columns that can be combined to form $\{\mathbf{v}_n\}$. A small value of $L-E$ and high $\epsilon$ lead to increased collision probability. As earlier mentioned, an increase in collision probability makes it difficult to identify the MTDs. It is therefore necessary to note that, to facilitate coexistence between eMBB and MTC, there are some key trade-offs that have to be made, particularly, by limiting the number of eMBB devices. 
\begin{figure}[t]
    \centering
    \includegraphics[width=0.45\textwidth]{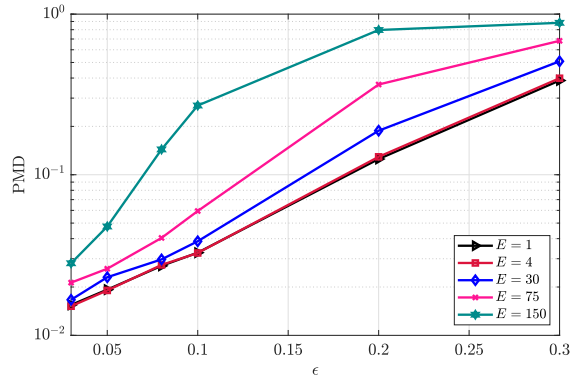}%\vspace{-2mm}
    \caption{PMD as a function of $\epsilon$ and $E$, $\mathrm{PFA} = 10^{-3}$}
    \label{htcVsMtd}%\vspace{-7mm}
\end{figure}}
%\subsubsection{Impact of eMBB pilot length}
%\subsubsection{Impact of Activation probability}
%\subsubsection{Impact of SNR levels}}
%%%%%%%%%%%%%%%%%%%%%%%%%%%%% 

\section{Conclusion and Future directions}\label{conclude}
We proposed a novel framework that enables the coexistence of heterogeneous services, thus, promising {for addressing the} spectrum scarcity problems in future cellular networks. Specifically, joint eMBB pilot and  sensing matrix design was {introduced}. {In this joint design, part of the sensing matrix is made orthogonal to the pilot sequences of the eMBB devices}, hence permitting training without interference. We found that our framework facilitates favourable coexistence with properly configured {average SNR} levels for both the eMBB devices and {the} MTDs. Similarly, the coexistence is also highly dependent on properly selecting the pilot length for the eMBB signals, while taking into consideration the activation probability {of the} MTDs.  Furthermore, our work revealed that increasing the number of antennas improves coexistence {of services}, which {constitutes a key} advantage of MIMO systems. Our work has, therefore, laid a foundation for a joint coherent and non-coherent signal processing framework, which can be of practical interest.  

In spite of the observed promising results, the applicability of the proposed solutions can be improved by deriving optimal points of operation in terms of pilot lengths and  average SNR from information theoretic perspectives. In addition, the present work assumes that active MTDs remain in that state for the whole coherence interval and as such assuming time-slotted activity. Future works can consider non-time-slotted (asynchronous) activity detection of MTDs in the coherence interval, thus improving the applicability of the proposed framework. Moreover, the current work can be extended to more general channel models such as Rician fading models.

\ifCLASSOPTIONcaptionsoff
  \newpage
\fi

\bibliographystyle{IEEEtran} 
%\bibliography{ref.bib}
%\begin{spacing}{1.31} %%COMPRESSION TWEAK
\bibliography{jour_short,conf_short,ref.bib}

% Generated by IEEEtran.bst, version: 1.14 (2015/08/26)
\begin{thebibliography}{10}
\providecommand{\url}[1]{#1}
\csname url@samestyle\endcsname
\providecommand{\newblock}{\relax}
\providecommand{\bibinfo}[2]{#2}
\providecommand{\BIBentrySTDinterwordspacing}{\spaceskip=0pt\relax}
\providecommand{\BIBentryALTinterwordstretchfactor}{4}
\providecommand{\BIBentryALTinterwordspacing}{\spaceskip=\fontdimen2\font plus
\BIBentryALTinterwordstretchfactor\fontdimen3\font minus
  \fontdimen4\font\relax}
\providecommand{\BIBforeignlanguage}[2]{{%
\expandafter\ifx\csname l@#1\endcsname\relax
\typeout{** WARNING: IEEEtran.bst: No hyphenation pattern has been}%
\typeout{** loaded for the language `#1'. Using the pattern for}%
\typeout{** the default language instead.}%
\else
\language=\csname l@#1\endcsname
\fi
#2}}
\providecommand{\BIBdecl}{\relax}
\BIBdecl

\bibitem{liu2018massive}
L.~Liu and W.~Yu, ``Massive connectivity with massive {MIMO}—{P}art {I}:
  Device activity detection and channel estimation,'' \emph{IEEE Transactions
  on Signal Processing}, vol.~66, no.~11, pp. 2933--2946, 2018.

\bibitem{liu2018massive2}
------, ``Massive connectivity with massive {MIMO}—{Part} {II}: Achievable
  rate characterization,'' \emph{IEEE Transactions on Signal Processing},
  vol.~66, no.~11, pp. 2947--2959, 2018.

\bibitem{mahmood2021machine}
N.~H. Mahmood \emph{et~al.}, ``Machine type communications: key drivers and
  enablers towards the {6G} era,'' \emph{EURASIP Journal on Wireless
  Communications and Networking}, vol. 2021, no.~1, pp. 1--25, 2021.

\bibitem{ghavimi2014m2m}
F.~Ghavimi and H.-H. Chen, ``{M2M} communications in {3GPP LTE/LTE-A} networks:
  Architectures, service requirements, challenges, and applications,''
  \emph{IEEE Communications Surveys \& Tutorials}, vol.~17, no.~2, pp.
  525--549, 2014.

\bibitem{zanella2014internet}
A.~Zanella \emph{et~al.}, ``Internet of things for smart cities,'' \emph{IEEE
  Internet of Things journal}, vol.~1, no.~1, pp. 22--32, 2014.

\bibitem{shariatmadari2015machine}
H.~Shariatmadari \emph{et~al.}, ``Machine-type communications: current status
  and future perspectives toward {5G} systems,'' \emph{IEEE Communications
  Magazine}, vol.~53, no.~9, pp. 10--17, 2015.

\bibitem{sharma2019toward}
S.~K. Sharma and X.~Wang, ``Toward massive machine type communications in
  ultra-dense cellular {IoT} networks: Current issues and machine
  learning-assisted solutions,'' \emph{IEEE Communications Surveys \&
  Tutorials}, vol.~22, no.~1, pp. 426--471, 2019.

\bibitem{senel2018grant}
K.~Senel and E.~G. Larsson, ``Grant-free massive {MTC}-enabled massive {MIMO}:
  A compressive sensing approach,'' \emph{IEEE Transactions on Communications},
  vol.~66, no.~12, pp. 6164--6175, 2018.

\bibitem{liu2018sparse}
L.~Liu \emph{et~al.}, ``Sparse signal processing for grant-free massive
  connectivity: A future paradigm for random access protocols in the {Internet
  of Things},'' \emph{IEEE Signal Processing Magazine}, vol.~35, no.~5, pp.
  88--99, 2018.

\bibitem{shahab2020grant}
M.~B. Shahab \emph{et~al.}, ``Grant-free non-orthogonal multiple access for
  {IoT}: A survey,'' \emph{IEEE Communications Surveys \& Tutorials}, vol.~22,
  no.~3, pp. 1805--1838, 2020.

\bibitem{popovski2019wireless}
P.~Popovski \emph{et~al.}, ``Wireless access in ultra-reliable low-latency
  communication {(URLLC)},'' \emph{IEEE Transactions on Communications},
  vol.~67, no.~8, pp. 5783--5801, 2019.

\bibitem{islam2016power}
S.~R. Islam \emph{et~al.}, ``Power-domain non-orthogonal multiple access
  {(NOMA)} in {5G} systems: Potentials and challenges,'' \emph{IEEE
  Communications Surveys \& Tutorials}, vol.~19, no.~2, pp. 721--742, 2016.

\bibitem{zhang2011sparse}
Z.~Zhang and B.~D. Rao, ``Sparse signal recovery with temporally correlated
  source vectors using sparse {Bayesian} learning,'' \emph{IEEE Journal of
  Selected Topics in Signal Processing}, vol.~5, no.~5, pp. 912--926, 2011.

\bibitem{huang2022noncoherent}
J.~Huang \emph{et~al.}, ``Noncoherent massive random access for inhomogeneous
  networks: From message passing to deep learning,'' \emph{IEEE Journal on
  Selected Areas in Communications}, vol.~40, no.~5, pp. 1457--1472, 2022.

\bibitem{Candes-Romberg-Tao-06}
E.~J. Cand\'{e}s, J.~Romberg, and T.~Tao, ``Robust uncertainty principles:
  Exact signal reconstruction from highly incomplete frequency information,''
  \emph{{IEEE} Trans. Inform. Theory}, vol.~52, no.~2, pp. 489--509, Feb. 2006.

\bibitem{donoho2006compressed}
D.~L. Donoho, ``Compressed sensing,'' \emph{IEEE Transactions on information
  theory}, vol.~52, no.~4, pp. 1289--1306, 2006.

\bibitem{Haupt-Nowak-06}
J.~Haupt and R.~Nowak, ``Signal reconstruction from noisy random projections,''
  \emph{{IEEE} Trans. Inform. Theory}, vol.~52, no.~9, pp. 4036--4048, Sep.
  2006.

\bibitem{lee2017packet}
B.~Lee \emph{et~al.}, ``Packet structure and receiver design for low latency
  wireless communications with ultra-short packets,'' \emph{IEEE Transactions
  on Communications}, vol.~66, no.~2, pp. 796--807, 2017.

\bibitem{lancho2021joint}
A.~Lancho, J.~{\"O}stman, and G.~Durisi, ``On joint detection and decoding in
  short-packet communications,'' \emph{arXiv preprint arXiv:2109.13669}, 2021.

\bibitem{ahn2019ep}
J.~Ahn, B.~Shim, and K.~B. Lee, ``{EP}-based joint active user detection and
  channel estimation for massive machine-type communications,'' \emph{IEEE
  Transactions on Communications}, vol.~67, no.~7, pp. 5178--5189, 2019.

\bibitem{wei2018joint}
Z.~Wei, D.~W.~K. Ng, and J.~Yuan, ``Joint pilot and payload power control for
  uplink {MIMO-NOMA} with {MRC-SIC} receivers,'' \emph{IEEE Communications
  Letters}, vol.~22, no.~4, pp. 692--695, 2018.

\bibitem{chen2019covariance}
Z.~Chen \emph{et~al.}, ``Covariance based joint activity and data detection for
  massive random access with massive {MIMO},'' in \emph{IEEE International
  Conference on Communications (ICC)}.\hskip 1em plus 0.5em minus 0.4em\relax
  IEEE, 2019, pp. 1--6.

\bibitem{huang2021compressed}
J.~Huang \emph{et~al.}, ``Compressed random access for noncoherent massive
  machine-type communications with energy modulation,'' \emph{IEEE Transactions
  on Wireless Communications}, 2021.

\bibitem{huang2020design}
------, ``Design of noncoherent communications: From statistical method to
  machine learning,'' \emph{IEEE Wireless Communications}, vol.~27, no.~1, pp.
  76--83, 2020.

\bibitem{chowdhury2016scaling}
M.~Chowdhury, A.~Manolakos, and A.~Goldsmith, ``Scaling laws for noncoherent
  energy-based communications in the {SIMO MAC},'' \emph{IEEE Transactions on
  Information Theory}, vol.~62, no.~4, pp. 1980--1992, 2016.

\bibitem{miao2016fundamentals}
G.~Miao \emph{et~al.}, \emph{Fundamentals of mobile data networks}.\hskip 1em
  plus 0.5em minus 0.4em\relax Cambridge University Press, 2016.

\bibitem{senel2018human}
K.~Senel, E.~Bj{\"o}rnson, and E.~G. Larsson, ``Human and machine type
  communications can coexist in uplink massive {MIMO} systems,'' in \emph{IEEE
  International Conference on Acoustics, Speech and Signal Processing
  (ICASSP)}.\hskip 1em plus 0.5em minus 0.4em\relax IEEE, 2018, pp. 6613--6617.

\bibitem{choi2017compressed}
J.~W. Choi \emph{et~al.}, ``Compressed sensing for wireless communications:
  Useful tips and tricks,'' \emph{IEEE Communications Surveys \& Tutorials},
  vol.~19, no.~3, pp. 1527--1550, 2017.

\bibitem{di2021dynamic}
R.~B. Di~Renna and R.~C. de~Lamare, ``Dynamic message scheduling based on
  activity-aware residual belief propagation for asynchronous {mMTC},''
  \emph{IEEE Wireless Communications Letters}, 2021.

\bibitem{di2020detection}
R.~B. Di~Renna \emph{et~al.}, ``Detection techniques for massive machine-type
  communications: Challenges and solutions,'' \emph{IEEE Access}, vol.~8, pp.
  180\,928--180\,954, 2020.

\bibitem{zhang2021unifying}
D.~Zhang \emph{et~al.}, ``Unifying message passing algorithms under the
  framework of constrained bethe free energy minimization,'' \emph{IEEE
  Transactions on Wireless Communications}, vol.~20, no.~7, pp. 4144--4158,
  2021.

\bibitem{kschischang2001factor}
F.~R. Kschischang, B.~J. Frey, and H.-A. Loeliger, ``Factor graphs and the
  sum-product algorithm,'' \emph{IEEE Transactions on information theory},
  vol.~47, no.~2, pp. 498--519, 2001.

\bibitem{wainwright2008graphical}
M.~J. Wainwright and M.~I. Jordan, \emph{Graphical models, exponential
  families, and variational inference}.\hskip 1em plus 0.5em minus 0.4em\relax
  Now Publishers Inc, 2008.

\bibitem{wang2018framework}
S.~Wang and N.~Rahnavard, ``A framework for clustered and skewed sparse signal
  recovery,'' \emph{IEEE Transactions on Signal Processing}, vol.~66, no.~15,
  pp. 3972--3986, 2018.

\bibitem{rush2018finite}
C.~Rush and R.~Venkataramanan, ``Finite sample analysis of approximate message
  passing algorithms,'' \emph{IEEE Transactions on Information Theory},
  vol.~64, no.~11, pp. 7264--7286, 2018.

\bibitem{takeuchi2020convolutional}
K.~Takeuchi, ``Convolutional approximate message-passing,'' \emph{IEEE Signal
  Processing Letters}, vol.~27, pp. 416--420, 2020.

\bibitem{rangan2019vector}
S.~Rangan, P.~Schniter, and A.~K. Fletcher, ``Vector approximate message
  passing,'' \emph{IEEE Transactions on Information Theory}, vol.~65, no.~10,
  pp. 6664--6684, 2019.

\bibitem{jiang2022performance}
J.-C. Jiang, H.-M. Wang, and H.~V. Poor, ``Performance analysis of joint active
  user detection and channel estimation for massive connectivity,'' \emph{IEEE
  Transactions on Signal Processing}, 2022.

\bibitem{tang2020device}
Z.~Tang \emph{et~al.}, ``Device activity detection and non-coherent information
  transmission for massive machine-type communications,'' \emph{IEEE Access},
  vol.~8, pp. 41\,452--41\,465, 2020.

\bibitem{djelouat2021joint}
H.~Djelouat \emph{et~al.}, ``Joint user identification and channel estimation
  via exploiting spatial channel covariance in {mMTC},'' \emph{IEEE Wireless
  Communications Letters}, 2021.

\bibitem{alsenwi2019embb}
M.~Alsenwi \emph{et~al.}, ``{eMBB-URLLC} resource slicing: A risk-sensitive
  approach,'' \emph{IEEE Communications Letters}, vol.~23, no.~4, pp. 740--743,
  2019.

\bibitem{abedin2018fog}
S.~F. Abedin \emph{et~al.}, ``Fog load balancing for massive machine type
  communications: A game and transport theoretic approach,'' \emph{IEEE
  Access}, vol.~7, pp. 4204--4218, 2018.

\bibitem{marata2021joint}
L.~Marata \emph{et~al.}, ``Joint channel estimation and device activity
  detection in heterogeneous networks,'' in \emph{29th European Signal
  Processing Conference (EUSIPCO)}.\hskip 1em plus 0.5em minus 0.4em\relax
  IEEE, 2021, pp. 836--840.

\bibitem{boyd2011distributed}
S.~Boyd, N.~Parikh, and E.~Chu, \emph{Distributed optimization and statistical
  learning via the alternating direction method of multipliers}.\hskip 1em plus
  0.5em minus 0.4em\relax Now Publishers Inc, 2011.

\bibitem{boyd2004convex}
S.~Boyd, S.~P. Boyd, and L.~Vandenberghe, \emph{Convex optimization}.\hskip 1em
  plus 0.5em minus 0.4em\relax Cambridge university press, 2004.

\bibitem{cai2011orthogonal}
T.~T. Cai and L.~Wang, ``Orthogonal matching pursuit for sparse signal recovery
  with noise,'' \emph{IEEE Transactions on Information theory}, vol.~57, no.~7,
  pp. 4680--4688, 2011.

\bibitem{bairagi2020coexistence}
A.~K. Bairagi \emph{et~al.}, ``Coexistence mechanism between {eMBB} and {uRLLC}
  in {5G} wireless networks,'' \emph{IEEE Transactions on Communications},
  vol.~69, no.~3, pp. 1736--1749, 2020.

\bibitem{ozdogan2019massive}
{\"O}.~{\"O}zdogan, E.~Bj{\"o}rnson, and E.~G. Larsson, ``Massive {MIMO} with
  spatially correlated {Rician} fading channels,'' \emph{IEEE Transactions on
  Communications}, vol.~67, no.~5, pp. 3234--3250, 2019.

\bibitem{albreem2021overview}
M.~A. Albreem \emph{et~al.}, ``Overview of precoding techniques for massive
  {MIMO},'' \emph{IEEE Access}, vol.~9, pp. 60\,764--60\,801, 2021.

\bibitem{ngo2020multi}
K.-H. Ngo \emph{et~al.}, ``Multi-user detection based on expectation
  propagation for the non-coherent {SIMO} multiple access channel,'' \emph{IEEE
  Transactions on Wireless Communications}, vol.~19, no.~9, pp. 6145--6161,
  2020.

\bibitem{wang2010unique}
M.~Wang, W.~Xu, and A.~Tang, ``A unique “nonnegative” solution to an
  underdetermined system: From vectors to matrices,'' \emph{IEEE Transactions
  on Signal Processing}, vol.~59, no.~3, pp. 1007--1016, 2010.

\bibitem{zheng2021jointMMV}
B.~Zheng \emph{et~al.}, ``Joint sparse recovery for signals of spark-level
  sparsity and {MMV} tail-$\ell_{2,1}$ minimization,'' \emph{IEEE Signal
  Processing Letters}, vol.~28, pp. 1130--1134, 2021.

\bibitem{eldar2012compressed}
Y.~C. Eldar and G.~Kutyniok, \emph{Compressed sensing: theory and
  applications}.\hskip 1em plus 0.5em minus 0.4em\relax Cambridge university
  press, 2012.

\bibitem{chen2018sparse}
Z.~Chen, F.~Sohrabi, and W.~Yu, ``Sparse activity detection for massive
  connectivity,'' \emph{IEEE Transactions on Signal Processing}, vol.~66,
  no.~7, pp. 1890--1904, 2018.

\bibitem{djelouat2021spatial}
H.~Djelouat, M.~Leinonen, and M.~Juntti, ``Spatial correlation aware compressed
  sensing for user activity detection and channel estimation in massive
  {MTC},'' \emph{IEEE Transactions on Wireless Communications}, 2022.

\bibitem{wipf2004sparse}
D.~P. Wipf and B.~D. Rao, ``Sparse bayesian learning for basis selection,''
  \emph{IEEE Transactions on Signal processing}, vol.~52, no.~8, pp.
  2153--2164, 2004.

\bibitem{tropp2006algorithms}
J.~A. Tropp, A.~C. Gilbert, and M.~J. Strauss, ``Algorithms for simultaneous
  sparse approximation. {Part I}: Greedy pursuit,'' \emph{Signal processing},
  vol.~86, no.~3, pp. 572--588, 2006.

\bibitem{steffens2018compact}
C.~Steffens, M.~Pesavento, and M.~E. Pfetsch, ``A compact formulation for the
  $\ell_ {2, 1}$ mixed-norm minimization problem,'' \emph{IEEE Transactions on
  Signal Processing}, vol.~66, no.~6, pp. 1483--1497, 2018.

\bibitem{pati1993orthogonal}
Y.~C. Pati, R.~Rezaiifar, and P.~S. Krishnaprasad, ``Orthogonal matching
  pursuit: Recursive function approximation with applications to wavelet
  decomposition,'' in \emph{Proceedings of 27th Asilomar conference on signals,
  systems and computers}.\hskip 1em plus 0.5em minus 0.4em\relax IEEE, 1993,
  pp. 40--44.

\bibitem{xia2018outage}
B.~Xia \emph{et~al.}, ``Outage performance analysis for the advanced {SIC}
  receiver in wireless {NOMA} systems,'' \emph{IEEE Transactions on Vehicular
  Technology}, vol.~67, no.~7, pp. 6711--6715, 2018.

\end{thebibliography}
%\end{spacing}

\end{document}